\documentclass[12pt,letterpaper,amsmath, showpacs, amssymb]{iopart}

\usepackage{graphicx}

\usepackage{color}
\usepackage{soul,xcolor}
\definecolor{tempcolor}{rgb}{0.0,0.0,0.0}

\usepackage{subfigure}  

\expandafter\let\csname equation*\endcsname\relax
\expandafter\let\csname endequation*\endcsname\relax

\usepackage{amsmath}

\usepackage{braket}

\usepackage{ulem}

\usepackage{cancel}

\usepackage{lineno}

\usepackage{xcolor}

\usepackage{booktabs}

\usepackage{cite}

\def\longrightharpoonup{\relbar\joinrel\rightharpoonup}
\def\longleftharpoondown{\leftharpoondown\joinrel\relbar}

\def\longrightleftharpoons{
  \mathop{
    \vcenter{
      \hbox{
	\ooalign{
	  \raise1pt\hbox{$\longrightharpoonup\joinrel$}\crcr
	  \lower1pt\hbox{$\longleftharpoondown\joinrel$}
	}
      }
    }
  }
}

\begin{document} 

\title{Many-body physics with ultracold plasmas: Quenched randomness and localization}
\author{John Sous$^{1,2}$ and Edward  Grant$^{1,3}$}
\address{$^1$Department of Physics \& Astronomy, University of British Columbia, Vancouver, BC V6T 1Z3, Canada \\
$^2$Stewart Blusson Quantum Matter Institute, University of British Columbia, Vancouver, British Columbia, V6T 1Z4, Canada \\
$^3$Department of Chemistry, University of
British Columbia, Vancouver, BC V6T 1Z1, Canada}
\eads{\mailto{jsous@phas.ubc.ca}, \mailto{edgrant@chem.ubc.ca}}

\date{\today}

\begin{abstract}

The exploration of large-scale many-body phenomena in quantum materials has produced many important experimental discoveries, including novel states of entanglement, topology and quantum order as found for example in quantum spin ices, topological insulators and semimetals, complex magnets, and high-$T_c$ superconductors.  Yet, the sheer scale of solid-state systems and the difficulty of exercising exacting control of their quantum mechanical degrees of freedom limit the pace of rational progress in advancing the properties of these and other materials.  With extraordinary effort to counteract natural processes of dissipation, precisely engineered ultracold quantum simulators could point the way to exotic new materials.  Here, we look instead to the quantum mechanical character of the arrested state formed by a quenched ultracold molecular plasma.  This novel class of system arises spontaneously, without a deliberate engineering of interactions, and evolves naturally from state-specified initial conditions, to a long-lived final state of canonical density, in a process that conflicts with classical notions of plasma dissipation and neutral dissociation.  We take information from experimental observations to develop a conceptual argument that attempts to explain this state of arrested relaxation in terms of a minimal phenomenological model of randomly interacting dipoles of random energies.   This model of the plasma forms a starting point to describe its observed absence of relaxation in terms of many-body localization (MBL).  The large number of accessible Rydberg and excitonic states gives rise to an unconventional web of many-body interactions that vastly exceeds the complexity of MBL in a conventional few-level scheme.  This experimental platform thus opens an avenue for the coupling of dipoles in disordered environments that will demand the development of new theoretical tools.

\end{abstract}

\section{Introduction}

When a transform-limited laser pulse excites an ensemble of isolated small molecules to a superposition state, the system evolves periodically with a well-defined pattern of quantum beats \cite{Huber}.  But, a wavepacket similarly created in a {\color{tempcolor}{system of}} large molecule{\color{tempcolor}{s}} dissipates exponentially to a state of maximum entropy \cite{Leitner}.  The Eigenstate Thermalization Hypothesis (ETH) explains this difference by arguing that the unitary dynamics of a superposition formed in a dense manifold of states always yield thermal expectation values as a time average \cite{Deustch,Srednicki,Tasaki,Rigol_Olshanii,Eisert,ChoasRev}.  Experiments confirm this idea, finding ergodic dynamics in quantum systems as small as three transmon qubits \cite{Neil}.  Statistical intramolecular energy redistribution forms the bedrock of more than 50 years of success in unimolecular reaction rate theory \cite{Baer}.  

But, just as methods of coherent control can overcome ergodicity and direct particular pathways for the chemical transformation of molecules, theory has described conditions under which certain many-body systems lack intrinsic decoherence, preserve spatial order and localize energy in highly excited states \cite{Gornyi,BAA,nandkishoreRev,AletRev,SidRev,AbaninRev}.  In these systems, transport paths for energy flow destructively interfere, {\color{tempcolor}{constraining}} a quenched system of quantum states to a corner of {\color{tempcolor}{its}} phase space.  The eigenstates of a many-body localized (MBL) phase retain a memory of their  initial conditions for arbitrarily long times.  This possibility to form enduring quantum superpositions embedded in interacting many-body ensembles suggests a potential as a means of preserving quantum information \cite{YaoQI,CiracQI}.  

Experimental quantum systems that fail to thermalize thus command a great deal of interest. A small number of highly engineered examples focussed on the dynamics of ultracold atoms have confirmed the principle of MBL \cite{MBL_Hubbard,MBL1D,MBL2D_2,MBL2D_1,Bloch_PRL,MBLopensystem,Monroe_MBL,rubio2018probing,lukin2018probing}. These results invite an intriguing question: In systems with stronger interactions and stronger disorder, could a state of many-body localization arise naturally? 

MBL acts in a profound way to constrain transport processes in a quantum system.  In the MBL phase, any local operator, time-evolved with the Hamiltonian and averaged over time, $\langle\hat O \rangle$, becomes a local integral of motion (LIOM) \cite{LIOMSerbyn,LIOMHuse,LIOMAbanin,LIOMSard1,LIOMSard2}.  Following any perturbation in a state of MBL, local integrals of motion associated with $\hat O$ survive for long times.  The LIOM refer to conserved quantities, for instance, a local particle or energy density.  As a system quenches, emergent LIOM determine the propagation of these conserved quantities in the MBL phase.  This naturally raises the question, in the progress of a quench, can locally emergent conservation laws act to guide the propagation of a spatially evolving quantum system to form a global many-body localized state?
 
In an experimental test of this question, we have studied the dynamics of a state-selected Rydberg gas of nitric oxide as it evolves to form a plasma \cite{Morrison2008,Sadeghi:2011,Saquet2012,Sadeghi:2014}.  On a microsecond timescale, this plasma bifurcates, irreversibly disposing electron energy to a reservoir of mass transport \cite{MSW_bifur}.  This evidently quenches the system to form an ultracold, quasi-neutral plasma in a state of arrested relaxation, far from thermal equilibrium \cite{Haenel}.  Classical models for plasma rate processes that describe the initial stages of Rydberg gas avalanche and plasma dissipation fail to account for the long lifetime and low electron temperature of this system \cite{Sadeghi:2011,Saquet2011,HaenelCRE}.  

This anomalous stability signifies a cessation of classical energy transport, which invites the theoretical question:  Does the arrested relaxation of this system signify self-assembly to a state of many-body localization?  Here, we refer to experimental observations and make logical arguments to derive a minimal model for particle interactions that accounts for the onset of the arrested phase  \cite{SousPRL}.  Within this model, we argue for the emergence of a transient state of many-body localization that could well explain these very slow dynamics.

\section{Experimental evidence for a state of arrested relaxation}

The molecular ultracold plasma forms a state of arrested relaxation in a sequence of steps, that begins with the electron impact avalanche of a prolate Rydberg gas ellipsoid.  A fraction of Rydberg molecules form pairs spaced within a critical distance, $r_P$.  These interact, releasing Penning electrons.  Starting in the core of the Rydberg gas ellipsoid, electron-Rydberg collisions drive an electron-impact ionization avalanche.  Some fraction of Rydberg molecules relax, releasing energy that heats electrons.  The electron gas expands, exhausting energy by radially accelerating NO$^+$ ions.  The moving NO$^+$ ions efficiently exchange charge with stationary NO$^*$ Rydberg molecules.  This redistributes momentum, and causes velocity-matched spatially correlated ions, cold electrons and Rydberg molecules to accumulate in the wings of the ellipsoid.  

We see these dynamics directly in three-dimensional images of plasma bifurcation \cite{MSW_bifur,Haenel}.  Here, the acceleration of opposing plasma volumes disposes kinetic energy initially present in the electron gas.  This sequence of events quenches and compresses these volumes to form ultracold, correlated distributions of NO$^+$ ions, electrons and Rydberg molecules.  

We use selective field ionization (SFI) spectroscopy to resolve the evolution of the electron binding energy as a function of time and initial Rydberg gas density, $\rho_0$.  The {\color{tempcolor}{short-time}} dynamics of electron-impact avalanche fit well with coupled-rate-equation simulations that consider $\rho_0$ and $n_0$, the selected principal quantum number of the Rydberg gas.  

\begin{figure}[h!]
\begin{center}
\includegraphics[width= 1 \columnwidth]{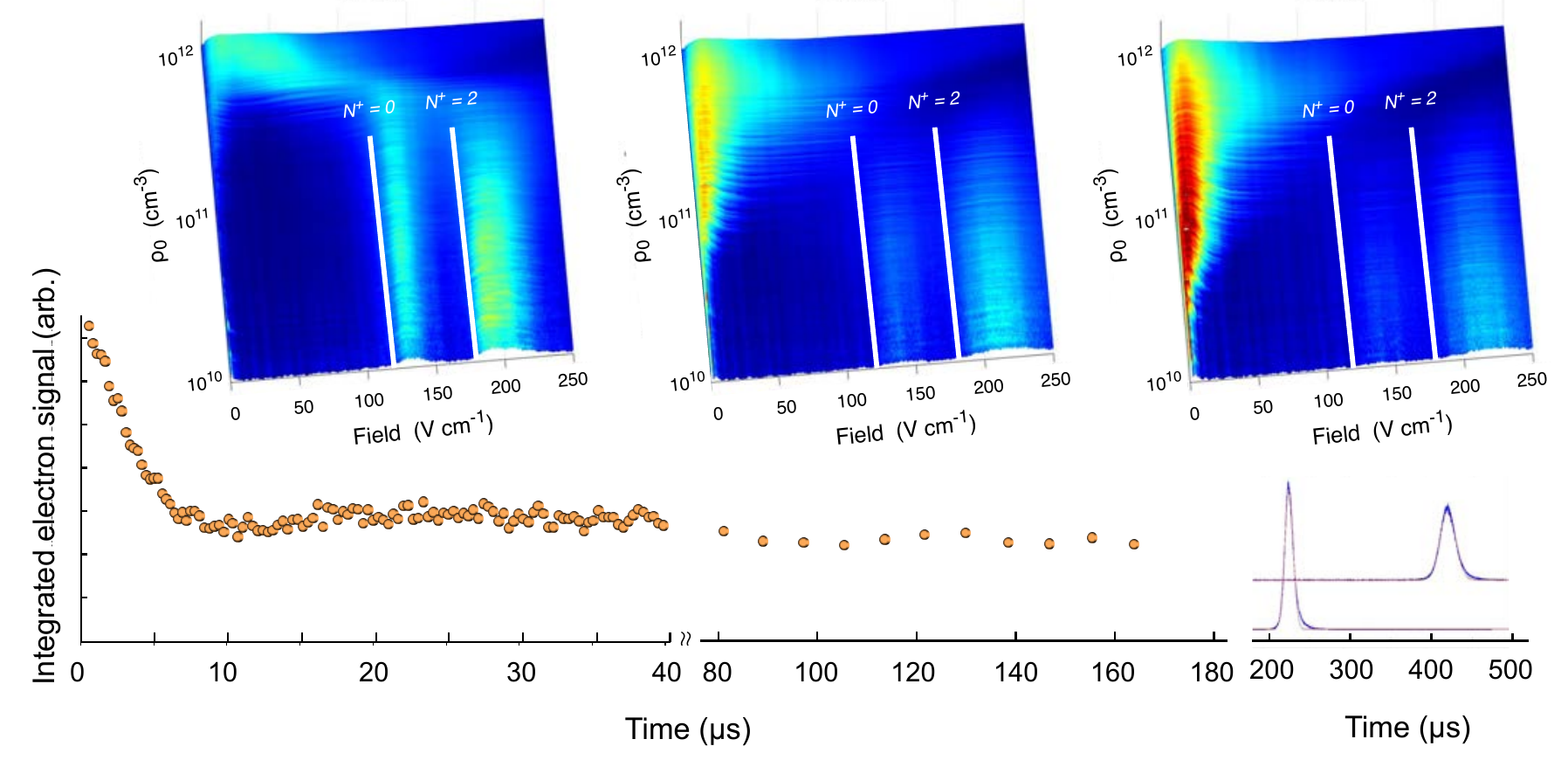}
\caption{top:  Selective field ionization (SFI) spectra of ellipsoidal $n_0 \ell(N^+)=44f(2)$ Rydberg gases of nitric oxide, obtained with fields that rise linearly at 0.6 V ns$^{-1}$, starting 0, 300 and 450 ns after excitation.  The figure for each ramp-field delay displays a stack of 4,000 traces ordered from top to bottom by decreasing initial Rydberg gas density.  The characteristic pair of features observed for Rydberg gases of lower density correspond to the field ionization of the single selected $44f(2)$ state to form ions with rotational quantum numbers of $N^+= 0$ and 2 respectively.  Note in particular how effects of electron-Rydberg collisions evident in shaping the distribution of Rydberg binding energies at high densities, and shifting the $l$-mixed $44f(2)$ signal to higher appearance potential between 0 and 300 ns, cease to matter in the evolution to 400 ns, while at lower density, $n=44$ features evolve to a state of very low binding energy without populating Rydberg states at intermediate energy.  Note also the clear V-shaped region {\color{tempcolor}excluding molecules of} very low binding energy, most evident for a ramp delay of delay of 450 ns.  This gap increases linearly with density {\color{tempcolor}reflecting} the $1/r^3$ variation in Rydberg-Rydberg dipolar coupling.  bottom:  Integrated electron signal as a function of ultracold plasma flight times from 0 to 400 $\mu$s. See reference \cite{Haenel}}  
\label{fig:n44SFI}
\end{center}
\end{figure}

Figure \ref{fig:n44SFI} shows a typical example.  SFI traces at lower initial density first exhibit the field ionization spectrum of a Rydberg gas in the initially selected $44f(2)$ state.  The spectrum then evolves owing to the effects of electron-Rydberg collisions.  Over a period of 300 ns, the spectrum directly shows evidence of electron-Rydberg collisional $l$-mixing, which shifts the signal produced at lower density by residual $n_0$ Rydberg molecules to a higher appearance potential.  On the same timescale, SFI traces show that the avalanche at higher density consumes all the $44f(2)$ Rydberg molecules, creating a narrow electron binding-energy distribution of width $W$, which peaks within 500 GHz of the zero-field ionization threshold, and tails to reflect a high-Rydberg population of intermediate $n$.  

After 300 ns, electron collisional redistribution of the Rydberg population over $n$ apparently stops.  The distribution of molecules with near-threshold binding energy rises while the residual population of $44f(2)$ molecules decays, and we see little or no change in the high-density distribution of intermediate-$n$ Rydberg molecules.  

As displayed in Figure \ref{fig:n44SFI}, the total amplitude of the detected electron signal falls with time from 0 to 5 $\mu$s.  This short-time decay, observed as a function of initial Rydberg gas density \cite{Sadeghi:2011}, accords with kinetic rate-theory models for charge neutralization by NO$^+ + {\rm e}^-$ dissociative recombination and NO$^*$ Rydberg predissociation \cite{Saquet2012, HaenelCRE}.  After 5 $\mu$s, neutral dissociation stops, and the ultracold plasma begins a phase of arrested relaxation that lasts for a time as long as we can make measurements.  While the rate processes that define this sequence of events depend on initial conditions, the system evolves to the same final state density and binding energy regardless of $\rho_0$ and $n_0$.

Experiment thus affords a direct view of a robust self-assembly in which a Rydberg gas undergoes a spontaneous avalanche to form a well-defined ensemble in which ultracold, correlated NO$^+$ ions and electrons occupy a binding energy distribution that extends 500 GHz below the ionization energy of nitric oxide.  Density-calibrated, time-resolved SFI spectra, such as those shown in Figure \ref{fig:n44SFI},  directly map the propagation of the electron-impact avalanche as it consumes the initially selected Rydberg gas. 

Images of this evolution projected in the $x,y$ plane perpendicular to the propagation axis, capture a cross-beam bifurcation that quenches the plasma to form separating volumes that expand very slowly \cite{MSW_bifur,Haenel}.  This slow rate of expansion tells us that free electrons, if present, must have an exceedingly low temperature.  But, a classical nitric oxide plasma {\color{tempcolor}{with high density and such a low electron temperature}} would be highly unstable:  Coupled rate-equation simulations predict the rapid evolution of such a system to an inevitable steady state of high electron temperature that proceeds with rapid expansion and dissipation to form neutral nitrogen and oxygen atoms \cite{Saquet2011,Saquet2012,HaenelCRE}.   

The long lifetime, low binding energy and slow rate of expansion might suggest a system quenched to a gas of very high-$n$ Rydberg molecules.  But again, for the conditions of these observations, such a Rydberg gas would be highly unstable with respect to classical collisional rate processes.  Coupled rate-equation simulations predict a rapid evolution essentially to the same steady state of high electron temperature and dissipation to neutral nitrogen and oxygen atoms \cite{HaenelCRE}.

Thus, the experiment captures definitive images of a plasma that has been quenched to a long-lived state that should be both thermodynamically unstable and kinetically unstable with respect to well-defined thermochemical rate processes.  We proceed now to develop a theoretical model for this state of arrested relaxation.  

\section{Theoretical rationale for the arrested state}

\subsection{Single-particle basis states}

From time-resolved SFI spectra, we find that the ellipsoidal nitric oxide Rydberg gas avalanches to a plasma, bifurcates, quenches, and compresses to form the same final state of ultracold NO$^+$ ions, electrons and Rydberg molecules, regardless of the initial principal quantum number or density.  Spatial correlation develops by virtue of Penning ionization coupled with dissociation during the avalanche to plasma \cite{Sadeghi:2014}, and momentum matching via charge exchange as the plasma undergoes bifurcation and quench.  We can assume that the combination of these effects gives rise to an emergent lattice of NO$^+$ charge centres \cite{PPR}. 

In an effort to describe this many-body state, let us start by adopting a simple basis consisting of the high-Rydberg states of NO together with an excitonic band of states formed by electrons weakly bound to ions in a dielectric medium containing a small excess of NO$^+$ ions.  Let us define this basis by the eigenfunctions of a Hamiltonian, $H_{\rm d}$:
\begin{equation}
{H_{\rm d}} = \sum_i \left ({\frac{{\bf P}_i^2}{2m}} + h_i \right)
\label{equ:ham1}
\end{equation}
obtained within the narrow binding-energy interval, $W$.  Within $H_d$, a local Hamiltonian, $h_i$, defines the eigenstates of each extravalent electron in the field determined by an NO$^+$ core in a dielectric medium defined by the surrounding ions and electrons.  

\begin{figure}[h!]
\begin{center}
\includegraphics[width= .5 \columnwidth]{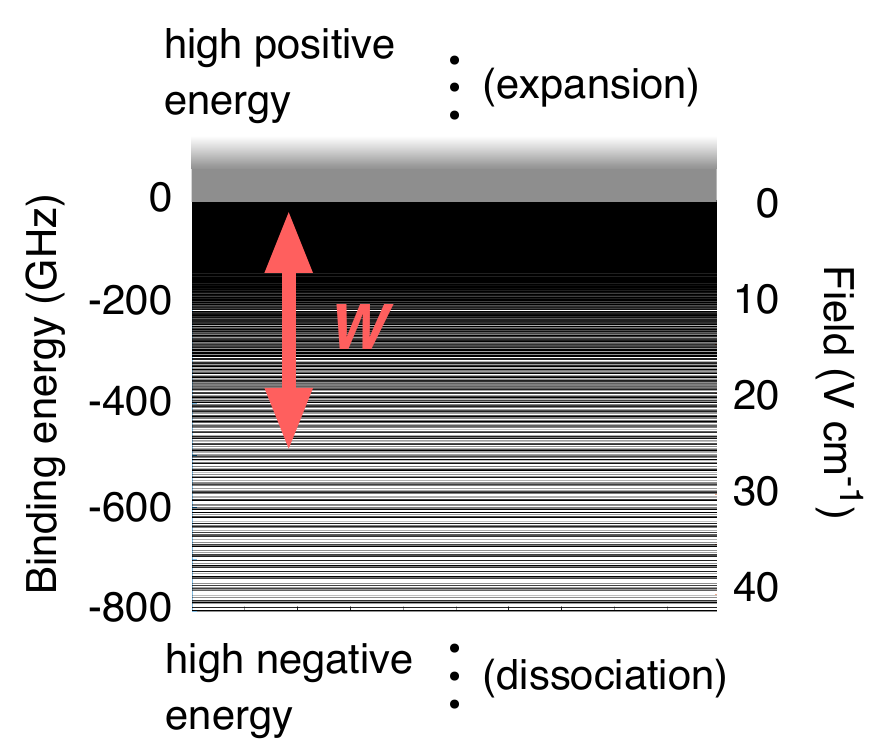}
\caption{Bifurcation and quench forms ions and electrons with measured distribution of binding energies, $W$.  The observations of little expansion after arrest bounds W from above, while the absence of dissociation after arrest bounds $W$ from below.  }
\label{fig:basis}
\end{center}
\end{figure}

For each dipole, $i$, $h_i$ is exceedingly complex.  Even for the limit of a completely isolated NO$^+$ and Rydberg electron, $h_i$ must account for electron orbital interactions with the rotational, vibrational and electronic degrees of freedom of the core ion, including in particular coupling to predissociative channels of irreversible decay to neutral products, N($^4$S) and O($^3$P)  \cite{Greene,Giusti}.  Extending $h_i$ to allow for excitonic binding adds considerable complexity \cite{Ramazanov,Stanton}.  But, nevertheless, we can expect its eigenstates to exhibit properties that vary smoothly as a function of binding energy \cite{Gallagher}, extending from a regime of isolated, central-field Rydberg molecules to one of Rydberg-like excitons.  Together, these form a basis of dipoles,  $\ket{e_i}$, with momenta ${\bf P}_i$.
\begin{equation}
\{ \ket{e_1}, \ket{e_2}, \ket{e_3} ... \}.
\label{dipole-basis}
\end{equation}

Figure \ref{fig:basis} schematically diagrams this basis.  Here we show the interval $W$ defined by SFI spectra and indicate a domain of lower energy, where more deeply bound Rydberg states predissociate on a microsecond timescale, and one of higher energy where faster-moving electrons drive substantial rates of plasma expansion.  Experimental observations of very long plasma lifetimes and very slow rates of plasma expansion exclude states in these domains from the basis we need to describe this system.

Bifurcation quenches the plasma to occupy levels in the interval, $W$, which for an average binding energy of 250 GHz, form a manifold with more than two states per MHz.  Over a period of 10 $\mu$s, the plasma evolves to an ellipsoidal arrested state with a spatially averaged density of $\sim 1.4 \times 10^{10}$ cm$^{-3}$, at which point a distance of 3.3 $\mu$m separates the average pair of Rydberg molecules \cite{MSW_tutorial,Haenel}.  The plasma quench distributes molecules randomly over states.  Rydberg-Rydberg interactions determine the properties of these states and the dynamics of energy transport with which the system continues to evolve.  We now consider these interactions and develop a model for the long-time dynamics.

\subsection{Intermolecular interactions}

Rydberg and Rydberg-like molecules represented by the states in our zeroth-order basis undergo interactions governed by the electrostatic potentials formed by NO$^+$ ion - electron pairs.  In a limit for which the intermolecular distance exceeds the dimensions of the individual Rydberg charge distributions, we can represent this by a multipole potential, ${{{{V}}}_{ij}} = {{{{V}}}}({\bf r_i - r_j})$, which, to lowest order reduces to the anisotropic dipolar interaction defined by  ${{V}^{\rm dd}_{ij}} = \left [{\bf d}_i \cdot  {\bf d}_j - 3({\bf d}_i\cdot {\bf r}_{ij})({\bf d}_j \cdot {\bf r}_{ij})\right ]  / {{ {\bf r}_{ij}^3}}$ \cite{Low}.  

In the quenched system, any pair of neighbouring dipoles $i$ and $j$  each occupy a random state of arbitrary energy, total angular momentum and parity.  Their dipole-dipole interaction encompasses hundreds if not thousands of paired, dipole-allowed transitions.  The strength of this interaction varies with the distance between NO$^+$ ions and the relative orientation of their internuclear axes.  Little of what follows depends upon the exact determination of matrix elements for particular interactions.  Where necessary for illustration, we shall average over solid angle in the molecular frame and estimate coupling strengths for partners separated by the average distance for a given density, as determined by the Wigner-Seitz radius, $a_{ws} =(3/4 \pi \rho)^{1/3}$.  Generally speaking, the amplitudes of individual dipole-dipole interactions increase with the average effective principal quantum number, approximately as $[(n_i + n_j)/2]^2$, { and fall off quadratically with the difference in principal quantum number} \cite{Samboy,Weber}.  A small subset of resonant transitions dominate the interaction of any particular pair of Rydberg/excitons.  

\subsubsection{Dipole-dipole Hamiltonian}

Taking ${H}_{\rm d}$ together with this interaction potential, the Hamiltonian up to pairwise dipole-dipole interactions is
\begin{equation}
{H} = \sum_i \left ({\frac{{\bf P}_i^2}{2m}} + h_i \right) + \sum_{i,j}{{{{V}^{\rm dd}_{ij}}}}.
\label{equ:ham}
\end{equation}
We can reduce this Hamiltonian to describe the pairwise interactions evaluated in the $\ket{e_i}$ basis \cite{Pupillo, Krems_book1}:
\begin{equation}
H^{\rm dd} =  \sum_i  {\frac{{\bf P}_i^2}{2m}} + \sum_{i,j}{{V}^{\rm dd}_{ij}}.
\label{equ:dham}
\end{equation}

Hamiltonians such as (\ref{equ:ham}) and (\ref{equ:dham}) most often refer to the case where a high-resolution laser prepares a Rydberg gas in which some dipole-dipole interaction gives rise to a particular set of coupled states \cite{Pillet,Low,Firstenberg}.  In the present case, avalanche and quench spontaneously form a molecular ultracold plasma in a process that populates a great many different states.  These states evolve spatially and interact without reference to a light-matter coherence or a dipole blockade of any kind

\subsubsection{Projection to a regime of lower-order resonance}

The experimental system relaxes in a few microseconds to a quenched state of ultracold temperature, from which it expands at no more than a few tens of meters per second.  Dipole-dipole coupling proceeds on an instantaneous timescale \cite{rydbergs1,rydbergs5,Barredo,rydberg_Gross}.  During the first 100 ns of the avalanche, under conditions well-described by classical coupled rate-equation kinetics, the SFI spectrum shows obvious effects of electron-Rydberg collisional $l$-mixing \cite{Haenel}.  Long-range NO$^+$ ion -- NO Rydberg collisions act to redistribute ion radial momentum and accelerate electron cooling.  Thereafter, as the system approaches a state of arrested relaxation, all signs of electron-Rydberg interaction disappear.  In particular, SFI spectra show no sign of electron-Rydberg $n$-changing collisions.  Heavy-particle, Rydberg-Rydberg inelastic collisions occur with many orders of magnitude lower frequency.  

During this latter phase of relaxation, we assume a regime of correlated e$^-$ - NO$^+$ exciton and NO* Rydberg motion, in which intermolecular potentials guide species to form a state of spatial correlation.  Here, dipole-dipole exchange dominates energy transport, evolving in an emergent, lattice-like frame of very slow residual relative motion of the ion centres \cite{Sadeghi:2014,MSW_tutorial}.
 
These well separated timescales allow us to write an effective Hamiltonian that describes pairwise interactions evolving in an approximate frame of stationary ions and Rydberg molecules: $H_{\rm eff} = \mathcal{P} \sum_{i,j}{{V}^{\rm dd}_{ij}}$, in which $\mathcal{P}$ represents a projector onto a Hilbert subspace characterized by low-order resonances.  In the plasma, no laser field acts to specify the particular states that couple in the interaction between any pair of dipoles, $i$ and $j$.  Solving the problem in full therefore requires a diagonalization of the complete dipolar matrix, encompassing all the states in a dense manifold that extends from $n \sim 80$ up to the weakest bound electronic state.  The subset of active levels varies widely from dipole to dipole in the plasma.  The use of $\mathcal{P}$ simplifies the task of conceptualizing this scenario by focusing attention on the comparably sparse set of most probable dipolar transitions in the arrest phase.  Note that this model keeps all low-order transitions, even those that involve interacting molecules with drastically different quantum numbers.  

\subsubsection{ Dipole-dipole coupling during the energy transport phase}

Let us consider an individual pair of dipoles in arbitrary states interacting under initial conditions of classical transport.  The dipole operator couples these molecules in an appropriate subspace spanned by the full set of accessible product states of the Rydberg single-particle wavefunctions.  This gives rise to a process of state mixing that we can represent as energy transfer by sequences of virtual one-photon exchanges.  Thus, a molecule of low principal quantum number can interact with one of high principal quantum number to exchange energy by instantaneous resonant interactions involving a total of $L$ states, $ \ket{e_1}, \ket{e_2}... \ket{e_L}$.  Likewise, two molecules that initially occupy states of similar principal quantum number couple in a comparable sequence, that transports one molecule to high $n$ and the other to low $n$ (where predissociation occurs with greater likelihood).  The experimental SFI spectrum in Figure \ref{fig:n44SFI} may well show direct evidence for such a process, in which for lower initial densities, two molecules with $n=44$ interact to transport one to a long-lived state with $n>80$ and the other to one to a dissociative level in the region of hydrogenic $n=34$.  

\subsubsection{Dipole-dipole coupling in the arrested phase}

The avalanche of the Rydberg gas to plasma, its hydrodynamic disposal of electron energy, its quench, correlation and transport to a regime of weak electron binding presages localization.  Interactions that occur in the transport phase alter the energy level structure of every molecule.  These perturbations that arise from disorder act to decrease or increase the degree of resonance in any particular coupling sequence.  { To occur with substantial effect, dipole-dipole interactions at any point in this process must have coupling amplitudes that exceed the energy deficit of transitions paired on $i$ and $j$.}  

The projector acts to limit the number of sequences in the scheme of dipole-dipole coupling permitted in the arrested phase.  The most extreme case, $L = 2$, considers only single-step transitions involving randomly selected basis states, $\ket{e_i^1}$, $\ket{e_i^2}$, $\ket{e_j^1}$ and $\ket{e_j^2}$. The case of $L=3$ couples three levels per molecule, in sequences of as many as two transitions.  For $L =4$, the active space increases to four basis states and includes the possibility of three transitions per molecule.  Each such calculation selects states, $1, 2 ... L$, randomly on each molecule, without bias.  It seems reasonable to expect conditions of resonance to limit the effectiveness of sequences with increasing order, but this remains to be determined.  Rare higher-order resonances could prove important.  

For now, we describe the arrested phase as an ensemble of random interacting dipoles in which pairwise interactions couple finite subsets of $L$-level basis states $\ket{e^1}$, $\ket{e^2}$, $ ... $ $\ket{e^L}$ that span the full space of all dipolar levels.  The superscript with lower (higher) integer label refers to the state with larger (smaller) electron binding energy.  Because these interactions occur between molecules quenched to random quantum states, the active levels, as they determine $L$ in each case, vary from one pair to the other.  Thus, our model imagines an effective dynamic quasi-resonant description with random dipolar matrix elements.  

The Rydberg-Rydberg coupling driven by ${V}^{\rm dd}_{ij}$ causes interacting states of nearly equal energy to repel.  The magnitude of the repulsion varies with the average intermolecular distance, as determined by sample density, and by the principal quantum number dependent average dipole-dipole coupling matrix element, $t$.  In the regime of energy transport, this level repulsion acts to produce a detectable void in the spectrum of levels populated in the dense system of states very near the ionization threshold, which is evident at high $\rho_0$ in the SFI spectrum (see Figure \ref{fig:n44SFI}).  The width of this void, as it varies with density, provides an experimental gauge of $t$, and from this, the average dipole-dipole coupling amplitude, $\tilde{J}$ at any given density.

Note that the projection we adopt here does not neglect Rydberg or excitonic dipoles of any particular binding energy.  On the contrary, $\mathcal{P}$ spans the full set of basis states, operating to pair all allowed low-order dipolar resonances.  Thus, dipoles in the arrested state of the ultracold molecular plasma dynamically couple via random dipolar interactions involving transitions in a $L$-level coupling schemes, where the $L$ levels change from dipole pair to dipole pair and from time to time in a random distribution from one pair of interacting dipoles to another.

\subsection{Effective many-body spin model}

\subsubsection{Spin Hamiltonian }
The quenched system supports a vast distribution of rare, resonant pair-wise interactions that occur in a random landscape of Rydberg/exciton energies.  Dipole-dipole coupling in this dense system of basis states causes excitation exchange.  Near-resonant interactions predominate. For simplicity, we assume these to involve $L$ states on each molecule, where $L$ is a small number in the range from 2 to 4.  These interactions form $L$-level systems drawing from {\it different } basis states from dipole to dipole: The states $\ket{e^1}$, $\ket{e^2}$ ... $\ket{e^L}$ vary from one dipole to another and from time to time. 

To describe this ensemble, we adopt the representation of a many-body Fock space of interacting dipoles  \cite{Fetter_and_Welcka}, representing pairwise excitations by spins with energies, $\epsilon_i$, and exchange interactions governed by an XY model Hamiltonian that describes these interactions in terms of their spin dynamics \cite{Gorshkov,XiangKrems}: 

\begin{equation} 
H_{\rm eff} =  \sum_i \epsilon_i \hat{S}^z_i + \sum_{i,j} J_{ij} (\hat{S}^+_i \hat{S}^-_j + h.c.),
\label{eqn:XY}
\end{equation}
where $\hat{S}^\gamma$ in each case denotes a spin operator acting in the space of L active states, and $\gamma = x,y$ or $z$. $h.c.$ refers to Hermitian conjugate.  

We adopt this Hamiltonian as a minimal model of the system immediately after the quench.  It reflects the dominant contributions to the diagonal and off diagonal disorder created by the random variation in the level system from dipole to dipole, as projected to low $L$.  Later we {\color{tempcolor} comment on} the evolution of this picture with slow expansion of the NO$^+$ ion lattice, and the possible sampling of dipole-dipole interactions in a regime of higher-$L$.  In what follows, we shall set $\hbar=1$.  

\subsubsection{On-site local spin term }
For each individual dipole $i$ described by states $\ket{e_i^1}$, $\ket{e_i^2} \dotso \ket{e_i^L}$ at a specific experimental instance, we define a projection operator for $\ket{e_i}$ states: $\hat{\sigma}^{e^l}_i= \ket{e^l_i }\bra{e^l_i }$. We can then represent the $l= 1, 2 ... L$ levels of a dipole $i$, with an energy spacing in the interval between $\ket{e_i^1}$ and $\ket{e_i^L}$, by a one-body local spin operator $\epsilon_i \hat{S}^z_i$.  This defines an energy $ \epsilon_i^{m_S} m_S$, where $S=(L-1)/2$ is the magnitude of the spin vector  i.e. $S=1/2$ for $L=2$, $S=1$ for $L=3$, $S=3/2$ for $L=4$, and $m_S$ is its projection, which runs from $-S$ to $S$.  

Thus, for $L=2$, we have $S=(L-1)/2 = 1/2$ and $m_S \in [-1/2,1/2]$.  So $\epsilon_i \hat{S}^z_i$ operates on $\ket{e_i^1}$ and $\ket{e_i^2}$ to yield energies $- \epsilon_i/2$ and $ \epsilon_i/2$, respectively,  measured from a midpoint energy of 0.  For $L=3$, we have $S=(L-1)/2 = 1$ and $m_S \in [-1,0,1]$ forms three levels, $- \epsilon_i^{m_S=-1}$, 0 and $ \epsilon_i^{m_S=1}$.  If the magnitude of $\epsilon_i^{m_S=-1}$ equals that of $\epsilon_i^{m_S=+1}$, this energy spacing is simply $2 \epsilon_i$.  The case of $\epsilon_i^{m_S=-1} \ne \epsilon_i^{m_S=+1}$ defines a spacing that is asymmetric by some quantity, $\delta_i$.  Similarly, for $l=4$, we have $S=(L-1)/2 = 3/2$, and $m_S \in [-3/2,-1/2,1/2,3/2]$ yields levels, $-3 \epsilon_i^{m_S=-3/2}/2,~ - \epsilon_i^{m_S=-1/2}/2,~  \epsilon_i^{m_S=1/2}/2$ and $3 \epsilon_i^{m_S=3/2}/2$.  

\subsubsection{Nature of disorder in $\epsilon_i$ }

Quenching results in a strongly disordered landscape of Rydberg/exciton energies that varies randomly from dipole to dipole.  The time-resolved SFI spectrum affords an electron binding-energy distribution that describes the range over which these level separations vary, and this directly reflects the disorder in these on-site energies.  In spin language, our Hamiltonian term $\sum_i \epsilon_i \hat{S}^z_i$ represents a Gaussian-distributed random local field that spans a width, $W$, which experimentally defines the range of $\epsilon_i^{m_S}$. This field inherently encodes the quantum state randomness imposed by the dynamics of the initial avalanche to plasma.  The representative SFI spectrum in Figure \ref{fig:n44SFI} directly gauges $W \sim 500$ GHz by the range of electron binding energies observed in the quenched ultracold plasma.  Thus, we see that the first term in $H_{\rm eff}$ describes the diagonal disorder arising from the random on-site energy of any particular dipole owing to its randomly quenched quantum state and randomly sampled two-level transition.

\subsubsection{Resonance and spin-exchange }
In the coupled-state subspace of a pair of interacting dipoles, the dipole-dipole matrix elements describe a resonant transfer of excitation from $i$ to $j$ with weight $J_{ij}$ \cite{agranovich,Gorshkov,XiangKrems}.  We represent this excitation exchange between pairs of dipoles by lowering and raising operators in the $L$ subspace: $\hat{\sigma}^{-}_{i}$ and its Hermitian conjugate ($h.c.$) $\hat{\sigma}^{+}_{i}$. An excitation transfer between dipoles $i$ and $j$ therefore corresponds to a spin flip-flop interaction $J_{ij}  (\hat{S}^+_i \hat{S}^-_j + h.c.) \propto  J_{ij} (\hat{\sigma}^+_i \hat{\sigma}^-_j + h.c.)$.  We emphasize that this term provides for collisionless excitation exchange in the plasma by resonant and thus evidently rare spin flip-flop interactions.  This interaction occurs over a range of length scales dictated by the spatial distribution of resonant spins, but this is an energy-conservative process.  First-order spin flip-flops  redistribute excitations spatially, but do not change the global distribution over energy states. 

The XY Hamiltonian restricted to low $L$ serves adequately to describe the short-time dynamics of the nitric oxide ultracold plasma immediately following quench.  As we shall see later, it readily predicts a possible transient condition of MBL governed by pair resonances that confine the NO$^+$ - e$^-$ system to a narrow band of energy above a low principal quantum number regime of fast predissociation and below a high electron temperature regime of fast expansion.

\subsubsection{Construction of the XY interaction for different values of $L$}
Let us now consider particular diagrammatic forms defining the XY model for coupled two-level interactions within the level systems we represent by $L = 2$, 3 and 4.

\subsubsection*{$L = 2$ case} 

\begin{figure}[h!]
    \centering
        \includegraphics[width= .8 \columnwidth]{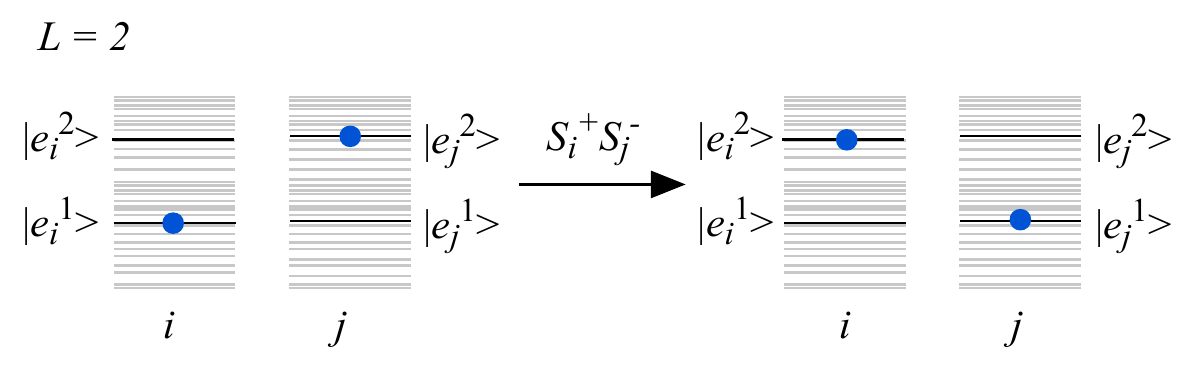}
    \caption {Energy levels of two Rydberg molecules, $i$ and $j$, dipole-coupled in the two-level approximation. The disordered environment of each molecule perturbs the energies states of $\ket{e_i}$ and $\ket{e_j}$.}
    \label{fig:N=2}
\end{figure}

For any given dipole $i$, defined for $L=2$ by states $\ket{e_i^1}$ and $\ket{e_i^2}$, we can write projection operators for the higher-energy state (spin-up) $\hat{\sigma}^{e^2}_i= \ket{e^2_i }\bra{e^2_i }= (1 + \hat{\sigma}_i^{z})/2$ and the lower-energy state (spin-down) $\hat{\sigma}^{e^1}_i = \ket{e^1_i} \bra{e^1_i} = (1 - \hat{\sigma}_i^{z})/2$.  We thus can represent the two levels of a dipole $i$ with spacing $\epsilon_i$, as in Figure \ref{fig:N=2}, in terms of a one-body operator $\epsilon_i \hat{S}^z_i = ( \epsilon_i / 2) \hat{\sigma}^z_i$, which yields an energy $\pm  \epsilon_i / 2$ depending on which state $\ket{e_i^2}$ or $\ket{e_i^1}$ is occupied, {\em i.e.} $\ket{e_i^2} \equiv \ket{\uparrow_i}$ and $\ket{e_i^1} \equiv \ket{\downarrow_i}$.  

Lowering and raising operators, $\hat{\sigma}^{-}_{i} = \ket{e^1_i} \bra{e^2_i}$ and its Hermitian conjugate $\hat{\sigma}^{+}_{i} = \ket{e^2_i} \bra{e^1_i}$, define a resonant spin flip-flop between dipoles $i$ and $j$: $J_{ij} (\hat{S}^+_i \hat{S}^-_j + h.c.) =  (J_{ij} / 2) (\hat{\sigma}^+_i \hat{\sigma}^-_j + h.c.)$ with amplitude $J_{ij} = {t_{ij}}/{{ r}_{ij} ^3}$, where $t_{ij} = \bra{e^2_i,e^1_j}V^{dd}_{i,j}\ket{e^1_i,e^2_j}$.  This refers to the dipole-dipole mediated transfer of excitation \cite{agranovich} represented by, for example, $\hat{S}^+_i \hat{S}^-_j \ket{\downarrow_i, \uparrow_j} = \ket{\uparrow_i ,\downarrow_j}$ {\em i.e.} $\ket{e_i^2,e_j^1} \xrightarrow[]{\hat{S}^+_i \hat{S}^-_j} \ket{e_i^1, e_j^2}$. We can expect this class of matrix element to be non-zero for many of the local eigenstates of $h_i$ and $h_j$, as the dipole-dipole operator couples states of opposite parity, limited only by selection rules governing the conservation of energy and total angular momentum \cite{BrownC}.

\subsubsection*{$L = 3$ case }
Dipole-dipole coupling schemes link sequences of transitions in cases of $L = 3$ and 4. Excitation transfer still governs the dynamics via terms like $J_{ij} (\hat{S}^+_i \hat{S}^-_j + h.c.)$, where the $\hat{S}$ operators live in the active $L$-dimensional subspaces.  

\begin{figure}[h!]
    \centering
        \includegraphics[width= .8 \columnwidth]{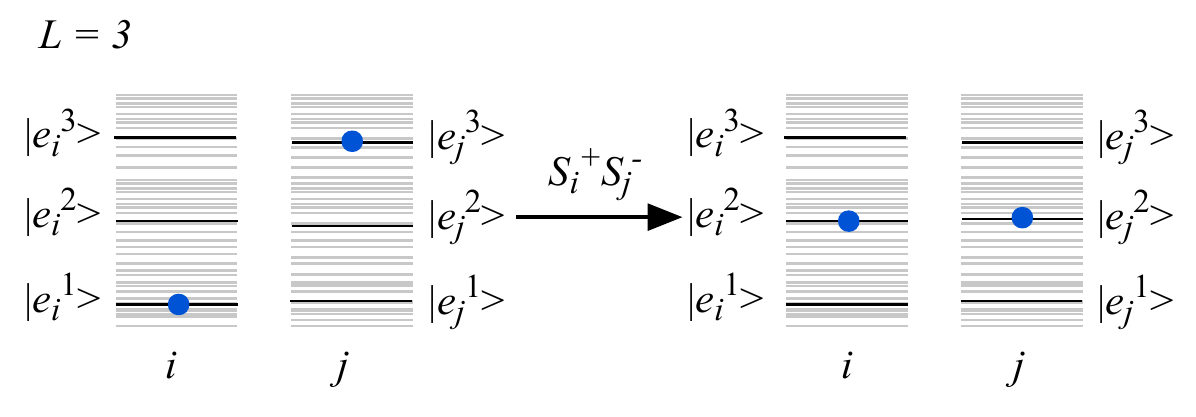}
    \caption {Schematic diagram representing a sequence in the coupling of two Rydberg molecules, $i$ and $j$, in a case of three active levels per molecule.}
   \label{fig:N=3}
\end{figure}

For example, a molecule initially in the random state $\ket{e_i^1}$ and one in the state $\ket{e_j^3}$ can interact via transitions involving a common final state, $\ket{e_i^2} = \ket{e_j^2}$, defining the sequence for $L=3$ diagrammed in Figure \ref{fig:N=3}.  

This type of three-state interaction figures prominently in studies of Rydberg quantum optics, in which a narrow bandwidth laser drives a resonant pair state.  Consider for example, the case of $58d_{3/2} + 58d_{3/2} \leftrightarrow 60p_{1/2} + 56f_{3/2}$ in $^{87}$Rb.  In zero field, at large separations these two-atom states of Rb exhibit a quasi-degeneracy, with an energy difference less than 7 MHz.  Using dipole traps to control the distance between pairs of $58d_{3/2}$ Rydberg atoms, Ga\"etan et al \cite{Gaetan} have showed that a separation of 4 $\mu$m breaks the degeneracy, forming a set of three eigenstates, mixed in the $p, d, f$ basis and separated by 50 MHz.  
{Excitation transfer operates in this case as:
\begin{equation} 
\label{two_spins}
\hat{S}^+_i \hat{S}^-_j \ket{S_i = -1, S_j = 1} = \ket{S_i = 0, S_j = 0}, 
\end{equation}
\noindent {\em i.e.} $\ket{e_i^1,e_j^3} \xrightarrow[]{\hat{S}^+_i \hat{S}^-_j} \ket{e_i^2, e_j^2}$.   
More generally, for $L=3$, a system randomly interacting dipoles couples to yield a mix of configurations, $\ket{e^2_i,e^2_j} \Longleftrightarrow \ket{e^3_i,e^1_j} \Longleftrightarrow \ket{e^1_i,e^3_j}$. }

\subsubsection*{$L = 4$ case }

\begin{figure}[h!]
    \centering
        \includegraphics[width= .8 \columnwidth]{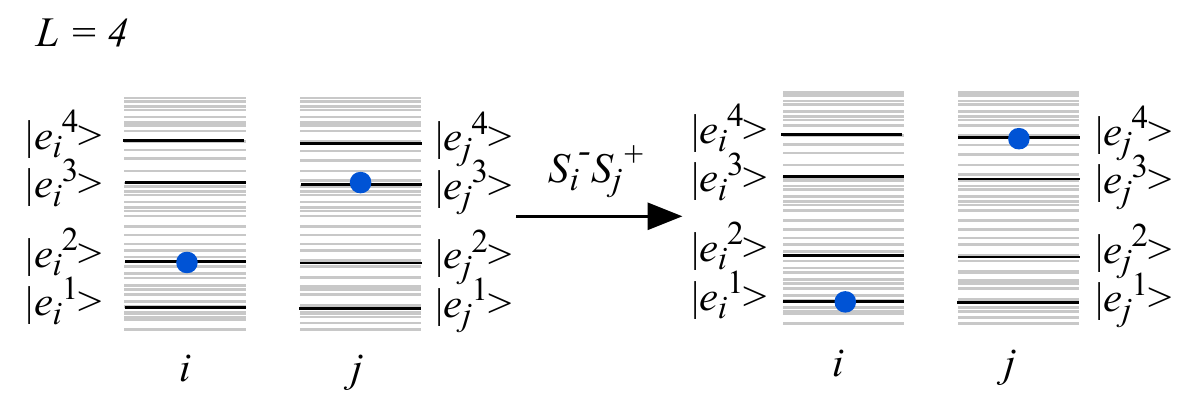}
    \caption {Schematic diagram representing a sequence in the coupling of two Rydberg molecules, $i$ and $j$, in a case of four active levels per molecule. The random population distribution over a very large number of Rydberg/exciton levels in the quenched ultracold plasma adds to the likelihood of $L=4$ and higher-order interactions.  Note that the state density increases rapidly with the elevation of population to higher principal quantum numbers.  This, accompanied by the growing rate of predissociation in the population of levels descending to $n$, gives the $L=4$ pathway a particular importance as the system relaxes to occupy a band of low binding energies during the transport phase that precedes arrest}
    \label{fig:N=4}
\end{figure}

For a disordered Rydberg gas occupying a dense manifold of states, the interaction described above as $L=3$ represents a special case of the more general $L=4$ coupling scheme diagrammed for particular levels of molecules $i$ and $j$ Figure \ref{fig:N=4}.  Here, the interaction of $i$ and $j$ give rise to eigenstates arising from the pair configurations, $\ket{e^2_i,e^3_j} \Longleftrightarrow \ket{e^3_i,e^2_j} \Longleftrightarrow \ket{e^1_i,e^4_j} \Longleftrightarrow \ket{e^4_i,e^1_j}$

{The particular interaction diagrammed above, which links a transition from $\ket{e_i^2}$ to $\ket{e_i^1}$ with the transition from $\ket{e_j^3}$ to $\ket{e_j^4}$, describes coupling in a subspace of $L=4$ with characteristics of the $S=1/2$ flip-flop interaction diagrammed in Figure \ref{fig:N=2}:  \begin{eqnarray} 
\label{three_spins}
\hat{S}^-_i \hat{S}^+_j \ket{S_i = -1/2, S_j = 1/2} = \ket{S_i  =  -3/2, S_j = 3/2}, 
\end{eqnarray}
with an excitation transfer: 
\noindent {\em i.e.} $\ket{e_i^2,e_j^3} \xrightarrow[]{\hat{S}^-_i \hat{S}^+_j} \ket{e_i^1, e_j^4}$. }

These partial coupling schemes link sets of flip-flop transitions on interacting dipoles.  A fuller treatment would apply the complete $S=1$ and 3/2 Pauli matrices to the cases of $L=3$ and 4, respectively, and extend similar sequences to higher $L$.  But, to a first approximation, we might assume that low-order resonant dipole-dipole excitation exchange in the dense manifold of basis states most prominently involves pairs of levels in these three low-$L$ cases. 

\subsubsection{Randomness in $J_{ij}$ }
$J_{ij} = {t_{ij}}/{ {r}_{ij}^3}$ determines the off-diagonal disordered amplitudes of the spin flip-flops.  To visualize the associated disorder, recognize that the second term varies as $t_{ij} \propto |{\textbf{d}_i}| |{\textbf{d}_j}|$, where every interaction selects a different ${\bf d}_i$ and ${\bf d}_j$.  The $C_3$ of 3230 MHz $\mu$m$^3$ measured by Ga\"etan et al. \cite{Gaetan} for the $58d_{3/2} + 58d_{3/2} \leftrightarrow 60p_{1/2} + 56f_{3/2}$ interaction in $^{87}$Rb scales as $n^2$ to yield a dipole-dipole coupling energy of 2 GHz at $n=146$ for a density 0.1 $\mu$m$^{-3}$.  This coupling width, representative of $\Delta n =2$ interactions at this binding energy, rises at high $n$ to span a large set of $|\epsilon_i-\epsilon_j|$ energy deficits.  Yet, this $J_{ij}$ is clearly much smaller than $W\approx 500$ GHz, the range of $\epsilon_i$, which is determined by the maximum separation between $\ket{e_i^1}$ and $\ket{e_i^L}$, as measured by the width of the binding energy distribution in the arrested state.  

\subsubsection{Induced Ising interactions }

\begin{figure}[h!]
    \centering
        \includegraphics[width= 1 \columnwidth]{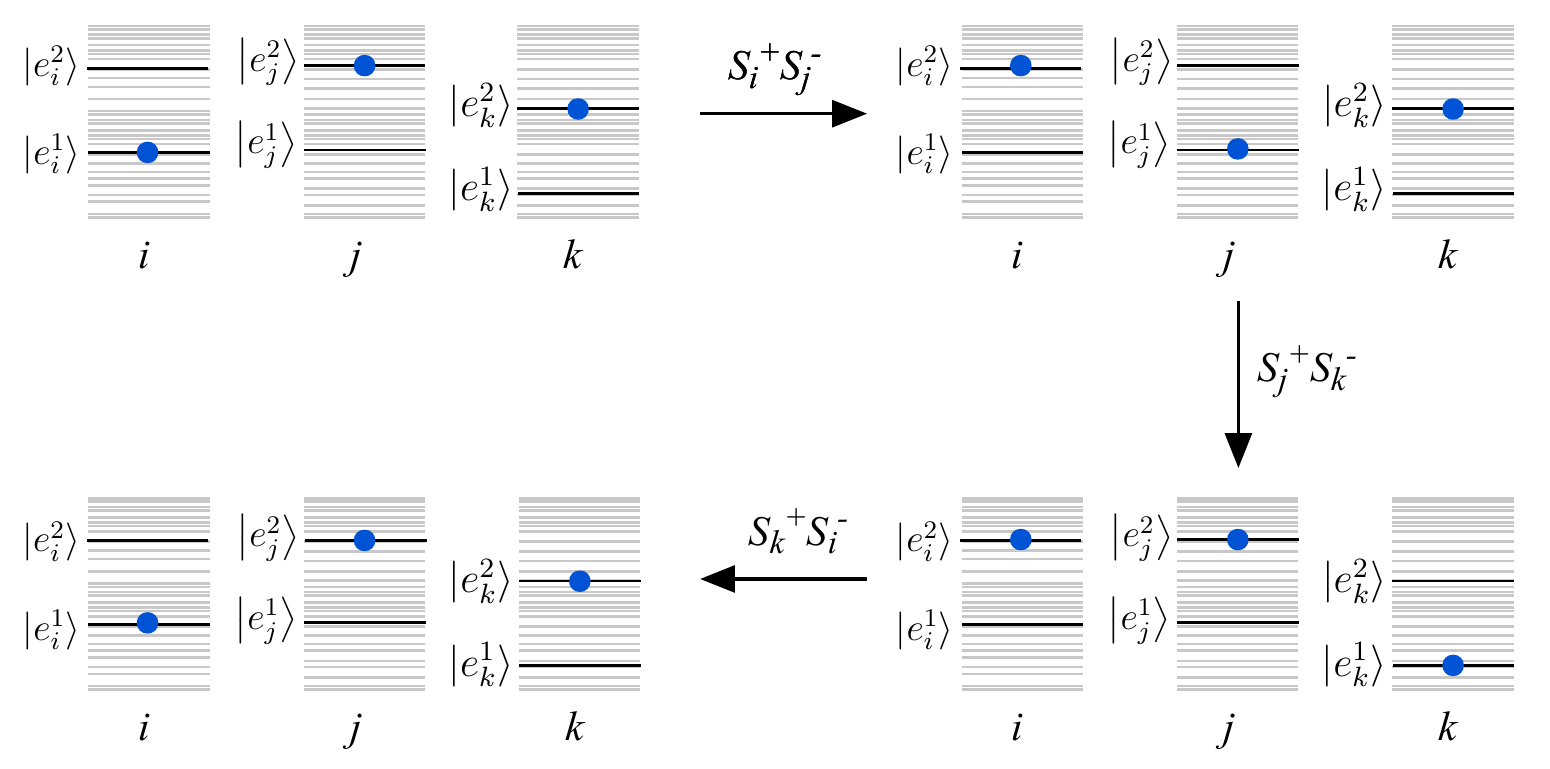}
    \caption {Schematic diagram representing three Rydberg molecules, $i$, $j$ and $k$, dipole coupled in a sequence of $L=2$ interactions that gives rise to a van der Waals term, $U_{ij} \hat{S}^z_i \hat{S}^z_j $ coupling $i$ and $j$ by a matrix element that varies as $1/r_{ij}^6$.  }
        \label{fig:vanderWaals}
\end{figure}

For this generic situation of $|J_{ij}| << W$, levels interact in sequence to add Ising terms that cause van der Waals shifts of pairs of dipoles \cite{Burin1}.   Figure \ref{fig:vanderWaals} diagrams an exemplary pairwise interaction of three nearest-neighbour spins $i$, $j$ and $k$ for the case of $L$ = 2.  Spins $i$ and $j$ couple via spin $k$ in a third-order process that proceeds as follows:  $\ket{\downarrow_i, \uparrow_j, \uparrow_k} \xrightarrow{\hat{S}^+_i \hat{S}^-_j} \ket{\uparrow_i, \downarrow_j, \uparrow_k} \xrightarrow{\hat{S}^+_j \hat{S}^-_k} \ket{\uparrow_i, \uparrow_j, \downarrow_k} \xrightarrow{\hat{S}^+_k \hat{S}^-_i} \ket{\downarrow_i, \uparrow_j, \uparrow_k}$, resulting in a self interaction that shifts the energies of $i$, $j$.  These processes occur with an amplitude, $U_{ij} \approx {J_{ij}^2\widetilde{J}}/{W^2}$, where $\widetilde{J}$  estimates $J_{ij}$, for an average value of $t_{ij}$ at an average distance separating spins \cite{Burin1}.  

Taken together with Eq (\ref{eqn:XY}) this result yields a general spin model with dipole-dipole and van der Waals interactions:
\begin{equation} \label{XY-Ising}
H_{\rm eff} = \sum_i \epsilon_i \hat{S}^z_i + \sum_{i,j} J_{ij} (\hat{S}^+_i \hat{S}^-_j + h.c.) + \sum_{i,j} U_{ij} \hat{S}^z_i \hat{S}^z_j,
\end{equation}
where $U_{ij} = {D_{ij}}/{r}_{ij}^6$ and $D_{ij} = {t_{ij}^2\widetilde{J}}/{W^2}$.  For $L>2$, Ising interactions still depend on distance as ${1}/{r}_{ij}^6$ , but fall linearly with disorder \cite{Burin1}. 

With the addition of this Ising term, the Hamiltonian develops many-body characteristics, which give rise to non-trivial dynamics involving emergent correlations between spins coupled by flip-flops.  For the simplest case of $L=2$, XY interactions cause dipoles, $i$ and $i'$, for which $S_i^z +S_{i'}^z = 0$, to entangle and form extended pairs when $\epsilon_i$ and $\epsilon_{i'}$ satisfy the resonance condition:
\begin{equation} 
\label{resonance}
\left | \epsilon_{i'} - \epsilon_i \right | < \left | J_{ii'} \right|
\end{equation}
where we understand the definition of $\epsilon_i $ and $ \epsilon_{i'}$ to account for perturbation by random fields that compare in magnitude to $\widetilde{J}$ \cite{Burin1}.  Thus, flip-flop interactions occur only for dipoles with nearly equal random fields.

Dipole-dipole coupling to form these resonant pairs dominates the short-range dynamics of the strongly disordered system formed by the quench of the ultracold plasma.  However, the Ising interaction links resonant pairs, $i,i'$, and $j,j'$ whenever $\epsilon_i - \epsilon_{i'} \approx \epsilon_j - \epsilon_{j'} $, even when $\epsilon_i$ differs a great deal from $\epsilon_j$.  Note that although the difference $\epsilon_i - \epsilon_{j}$ spans the full width of the disorder energy, differences in resonance widths will always be of the order $\widetilde{J}$, which satisfies the resonance condition for an Ising interaction for a fraction of pairs determined by $W$.  

These Ising interactions mix resonant pairs, creating a path for energy transport that becomes significant for a system of size $R$ when the average Ising interaction, ${\color{tempcolor}{U_{ave}(R)}}$, exceeds the average dipole-dipole pair interaction, {\color{tempcolor}$J_{ave}(R)$}.  When the coupling widths of extended-pair flip-flop interactions substantially exceed the Ising interaction width, the Ising interaction acts as an off-resonant perturbation that cannot lead to real energy transport between pairs. Variations on these dynamics of dipole-dipole and Ising interactions are likely to govern more complex states of entanglement encountered for cases of higher $L$, where the Ising interaction develops at second order in perturbation theory with less sensitivity to disorder \cite{Burin1}.

\section{Solution of the many-body Hamiltonian}
Even with its simplifying approximations, the complexity of the effective Hamiltonian places an exact solution of Eq (\ref{XY-Ising}) beyond reach for a plasma of ${10^8}$ molecules. The number of basis set wavefunctions needed for a full computation {\color{tempcolor}{that considers}} all possible initial spin state configurations scales as $L^{N_{\rm plasma}}$ or $L^{10^8}$.  Nonetheless we can gauge some likely properties of such a solution by analytical arguments.

\subsection{Locator expansion approach and resonance counting arguments}

To proceed this this respect, we start with an assumed state of complete localization and then ask whether delocalizing perturbations can destabilize this phase.  Here, we cast the problem in the form of the Hamiltonian, Eq (\ref{eqn:XY}), and expand about the infinite disorder limit where a localized phase must exist.  We ignore the disorder in $J_{ij}$, and focus on dipole-dipole interactions in the $L=2$ limit.  Thus, with reference to Eq (\ref{eqn:XY}), we develop a locator expansion in dipole-dipole perturbations, $H_{\rm D} = \sum_{i,j} J_{ij} ( \hat{S}^+_i \hat{S}^-_j + h.c.)$, about the localized state defined by $H_0 = \sum_i \epsilon_i \hat{S}^z_i $. 

We consider transport by dipole-dipole interactions along a chain of $n$ dipoles from a position, $\pi(0)=a$ to $\pi(n)=b$.  The resolvent \cite{Antonello_lecturenotes}:
\begin{equation} \label{resolvent}
G(b,a,\omega) = \sum_{\pi: a\rightarrow b, |\pi|=n}   \prod_{s=1}^n J_{\pi(s)}  \frac{1}{\omega - \epsilon_{\pi(s)} - \Sigma_{\pi(s)}^{\{ \pi(0),\dotso,\pi(s-1) \}}},
\end{equation}
propagates the wavefunction to point {\color{tempcolor}{$b$}}, given its properties at {\color{tempcolor}{$a$}}.  Here, $\sum_{\pi: a\rightarrow b, |\pi|=n}$ sums over paths on the web of dipoles from $a$ to $b$, and $\Sigma_{\pi(s)}^{\{ \pi(0),\dotso,\pi(s-1) \}}$ denotes the self-energy, having incorporated the contributions of different delocalizing paths to $\pi(s)$ and removed the points $\pi(0),\dotso,\pi(s-1)$. This procedure involves no double counting of paths. The self-energies converge to $0$ in the infinite disorder limit as propagation of excitations completely breaks down. For more details, see reference \cite{Antonello_lecturenotes}.

The expansion defined by the resolvent accounts for all possible pathways for the propagation of excitations.  Delocalizing interactions, which in effect increase this number to infinity, cause $G(b,a,\omega)$ to diverge.  The complexity of the present problem prevents a complete computation of the resolvent.  As an alternative, we can gauge the divergence of $G(b,a,\omega)$ by counting the number of resonances defined by a first-order perturbative condition in which the dipole-dipole coupling between two distant sites $i$ and $j$, $ J_{ij}=t_{ij}/r_{ij}^3$, substantially exceeds the energy deficit of the coupled transition ($ |J_{ij}| /|\epsilon_i - \epsilon_j| >> 1$).  A number of resonances as a function of system size that diverges in the thermodynamic limit indicates a breakdown of the locator expansion.  Such divergence would suggest an infinite number of pathways for excitations to propagate in the system, and thus delocalization on some finite timescale.  In the development that follows, we consider the resonance structure that develops for the specific case of dipole-dipole interactions in a three-dimensional system.

\subsection{Single-particle problem }

In the limit of a single excitation, the Ising interaction has zero amplitude, and the problem reduces to the single-particle dipolar XY Hamiltonian.  According to locator-expansion arguments, dipolar interactions give rise to extended states with subdiffusive dynamics in three dimensions \cite{Anderson, Levitov}.  Recent results show that while dipolar interactions in three dimensions can lead to extended states, such systems exhibit non-ergodic behviour \cite{Deng_nonergodic}.  Moreover, algebraic localization (as opposed to exponential Anderson localization) arises in systems with off-diagonal disorder in the long-range spin flip-flop interactions in one dimension \cite{Deng_algebraic}.  Thus, in a model like Eq. (8), with a mixture of diagonal and off-diagonal disorder, we might expect to find some form of localization in the single-excitation limit.

\subsection{Many-body problem }

For a system of many coupled excitations, Ising interactions form off-diagonal matrix elements in the resonantly coupled pair states \cite{Burin2}, which allow energy to propagate from one pair to the other. The progression of these interactions {\color{tempcolor}{leads}} ultimately to delocalization.  

In the development that follows, we utilize the treatments of references \cite{Yao, Burin1} to explore the particular consequences of dipole-dipole coupling in the three-dimensional system described by Eq  (\ref{XY-Ising}) for the case of $L=2$.  We represent typical values of $t_{ij}$ and $D_{ij}$ by $t$ and $D$ respectively, and refer again to an average dipole-dipole coupling term, $\widetilde{J}$, which constrains this picture to a limit without off-diagonal disorder.  To start, we simply explore the effect of the $xy$ spin flip-flops, and then consider the additional effects of $z$ Ising interactions on these couplings, as sketched below.

\subsubsection{Spin flip-flop interactions }
Spin flip-flop interaction proceeds via dipole-dipole matrix elements with amplitudes that exceed the energy deficit {\color{tempcolor}{of}} the corresponding transition: $t_{ij} / |r^3_{ij}| \geq \delta_{ij} = |\epsilon_i - \epsilon_j|$, where $\epsilon_i$ and $\epsilon_j$ span the measured interval of energy, $W$.  We consider a spherical shell surrounding each dipole that extends from a radial distance $R_{xy}$ to $2R_{xy}$.  On average, a central dipole sees a number of resonant spins for $R_{xy}<|r_{ij}|<2R_{xy}$ defined by:
\begin{equation} \label{LE1}
N_{xy}(R_{xy}) \sim\  ({\rho} R_{xy}^3) \frac{{t}/R_{xy}^3}{W} 
\end{equation} 
\noindent or,
\begin{equation}
N_{xy}   = {\rho} t/W,
\end{equation}
where $\rho$ is the density of spins (dipoles) in the plasma and $t$ represents an average coupling amplitude.  Note, because the number of particles in a shell increases as $R_{xy}^3$ while the resonance width decreases as $R_{xy}^{-3}$,  dipole-dipole coupling in three dimensions forms a critical case in which, for a given density, $ {\rho}$, the number of resonant dipoles, $N_{xy}$ does not change with radial distance.  

\subsubsection{Ising interactions} 
Resonant pairs, mixed by flip-flop interactions, define a new degree of freedom, termed a pseudospin, with an average energy splitting, $J(R_{xy}) =t/R_{xy}^3$.  Ising interactions couple pairs of pseudospins     {\color{tempcolor} by a term, $U(R_{z}) = D/R_{z}^6$, where $R_z$ {refers to}} a distance between pseudospins and $D$ again represents an average value of the van der Waals matrix element defined by Eq (\ref{XY-Ising}).  

The quantity $\varrho(R_{xy}) = \rho N_{xy}(R_{xy})$ defines the density of pseudospins of radial size, $R_{xy}$.  For such a pseudospin, we can determine a number of Ising resonances present within a shell from distance $R_{z}$ to distance $2R_{z}$ by,
\begin{equation} \label{LE2}
N_{z}(R_{xy},R_{z}) \sim [\varrho(R_{xy})R_{z}^3] \frac{D/R_{z}^6}{t/R_{xy}^3}  = \rho^2 \left( \frac{t}{W} R_z^3\right) \frac{D/R_{z}^6}{t/R_{xy}^3}.
\end{equation}

If $N_z(R_{xy},R_z)$ diverges as $R_z \rightarrow \infty$, the corresponding system delocalizes owing to a diverging number of energy transfer pathways.  To apply Eq (\ref{LE2}) in a test for this, we must know the typical pseudospin size.  As pointed out by Yao and coworkers \cite{Yao}, the effective size of a central pseudospin, $R_{xy}$, ought to grow with $R_z$.  Enforcing an extended-pairs condition for this case yields $R_{xy} \sim (t/D)^{1/3} R_{z}^{2}$ \cite{Yao}, which determines a number of Ising resonances that grows with distance as:
\begin{equation}  \label{LE3}
N_{z}(R_{z}) \sim \rho^2 \frac{t}{W} R_z^3.
\end{equation}
This quantity diverges like the volume as we take $R_z \rightarrow \infty$.  

\subsubsection{Critical system size }

As described above, for a three-dimensional system of fixed density spin flip-flop interactions {\color{tempcolor} entangle} dipoles to create a number of pseudospins that is critical in the system size.  In other words, we find a number of pseudospins that is invariant with coupling distance.  Ising interactions couple these pseudospins, mixing and energetically displacing states in the pseudospin basis.  The number of Ising resonances grows as the third power of the interaction distance $R$.   For conditions under which {\color{tempcolor}the} strength of these interactions exceeds the coupling energy of the pseudospins, they create a web of dynamic coupling that leads to delocalization \cite{Burin1}.   

The average energy splitting of resonant pairs coupled by spin flip-flop interactions is {\color{tempcolor}{$J_{ave}(R) = t/R^3$}} where we define the average coupling term by $t = \widetilde{J}/ \rho$.  The density of these resonant pairs with any size $R$ is simply, 
\begin{equation} \label{Nc1} 
\varrho(R) \sim \rho N_{xy} = \rho \Big( \frac{\widetilde{J}}{W} \Big).
\end{equation}
Defining an average Ising interaction $D = \widetilde{U}/\rho^{{6/3}} = \widetilde{U}/\rho^{{2}}$, we compute the Ising interacting energy density between pseudospins as
\begin{equation} \label{Nc2} 
{\color{tempcolor}{U_{ave}(R) \sim\ D  \varrho(R)^{2}  = \widetilde{U}  \left( \frac{\widetilde{J}}{W} \right)^2}} .
\end{equation}

Generally speaking dipole-dipole coupling is stronger in first order than second order.  But, for the system of pseudospin pairs, we see above that the quantity ${\color{tempcolor}{U_{ave}(R)}}$, which gauges the delocalizing effect of Ising interactions, decreases with $R$ more slowly than ${\color{tempcolor}{J_{ave}(R)}}$, the energy that couples dipoles to form the average pseudospin.  

So, in a system of finite size, there must exist a range $R_c$, that encompasses a critical number of dipoles, $N_c$, above which the coupling energy of the dipole pairs falls below the strength of the Ising interactions, allowing the system to delocalize \cite{Burin1}.  To determine $N_c$ we compare ${\color{tempcolor}{U_{ave}(R)}}$, the average Ising interaction energy coupling extended pseudospins to ${\color{tempcolor}{J_{ave}(R)}}$, the energy on average that couples the dipoles to form pseudospins.  The value, $R_c$, at which the two energies match determines a maximum system size for which delocalization {\color{tempcolor} prevails} for a given strength of disorder $W$.  Thus, to find $R_c$, we solve ${\color{tempcolor}{U_{ave}(R_c)}} = {\color{tempcolor}{J_{ave}(R_c)}}$ to obtain:

\begin{equation} \label{Nc3} 
R_c = \rho^{{-1/3}} \left( \frac{\widetilde{J}}{\widetilde{U}}\right)^{1/3} \left( \frac{{W}}{\widetilde{J}}\right)^{2/3}
\end{equation}
and the resulting condition is
\begin{equation} \label{Nc4} 
N_c = \rho R_c^3 = \left ( \frac{\widetilde{J}}{\widetilde{U}}\right) \left( \frac{{W}}{\widetilde{J}}\right)^2 = \left( \frac{{W}}{\widetilde{J}}\right)^4,
\end{equation}
where we used the definition $D_{ij} = {t_{ij}^2\widetilde{J}}/{W^2}$ to estimate $\widetilde{U} = {{\widetilde{J}}^3}/{W^2}$ in the last step.

\subsubsection{An estimate of $N_c$ for the experimental system  } \label{computeNc}

Let us compare this theoretical estimate with an experimental determination of the number of dipoles present in the quenched state of the ultracold plasma.  With precise control of the laser-crossed molecular beam geometry, we can fix the distribution of nitric oxide density in the volume illuminated to form an initial Rydberg gas \cite{MSW_tutorial}.  Generally speaking, $\omega_2$ saturates the second step of double resonance, so a varied intensity of $\omega_1$ determines the peak density of the Rydberg gas volume up to a maximum of $6 \times 10^{12}~{\rm cm}^{-3}$ ($6~ \mu{\rm m}^{-3}$).  This Rydberg gas density fluctuates, but we have managed to calibrate a measure of the sum of counts in each SFI trace to classify the corresponding initial density of every shot.  Coupled rate simulations that describe the kinetics of the avalanche of Rydberg gas to plasma conform well with the observed evolution in electron binding energy as a function of Rydberg gas density estimated in this way \cite{Haenel,HaenelCRE}.  

\begin{table}[h!]
 \caption{ Radial distribution of ion density in a Gaussian ellipsoid shell model for a quenched ultracold plasma of NO as it enters the arrest state with a peak density of $4 \times 10^{10}~{\rm cm}^{-3}$  ($0.04~ \mu{\rm m}^{-3}$), $\sigma_x = 1.0$ mm, $\sigma_y = 0.55$ mm and $\sigma_z = 0.70$ mm.  Here, the quasi-neutral plasma contains a combined total of  $1.9 \times 10^{8}$ NO$^+$ ions and NO Rydberg molecules.  The average density of this plasma is $1.4 \times 10^{10}~{\rm cm}^{-3}$ and the mean distance between NO$^+$ ions is 3.32 $\mu$m.} 
 \label{tab:ellipse}
  \begin{center}
 \begin{tabular}{rlccrc}  
\toprule
  \hspace{3 pt}   Shell  &  \hspace{2 pt}  Density &   Volume   &  \hspace{2 pt} Particle &    \hspace{2 pt}  Fraction  &   $a_{ws}$  \\  
Num &   \hspace{9 pt}  cm$^{-3}$ &    cm$^{3}$  & \hspace{2 pt} Number & $\times 100$ \hspace{5 pt} &  $\mu$m \\
 \hline 
    $1$ \hspace{5 pt} & $4.0 \times 10 ^{10}$  \hspace{3 pt} & $1.8 \times 10 ^{-6}$ &  \hspace{5 pt}  $7.0 \times 10 ^{4}$ &  $0.04$ \hspace{4 pt} & $1.81$  \\ 
   $2$  \hspace{5 pt} & $3.9 \times 10 ^{10}$ & $1.1 \times 10 ^{-5}$ &  \hspace{3 pt}  $4.4 \times 10 ^{5}$ & $0.23$  \hspace{4 pt} & $1.83$  \\ 
   $3$  \hspace{5 pt} & $3.7 \times 10 ^{10}$ & $9.0 \times 10 ^{-5}$ &  \hspace{3 pt}  $3.3 \times 10 ^{6}$ & $1.75$  \hspace{4 pt} & $1.86$  \\ 
   $4$  \hspace{5 pt} & $3.3 \times 10 ^{10}$ & $2.8 \times 10 ^{-4}$ &  \hspace{3 pt}  $9.3 \times 10 ^{6}$ & $4.87$  \hspace{4 pt} & $1.93$  \\ 
   $5$  \hspace{5 pt} & $2.6 \times 10 ^{10}$ & $8.3 \times 10 ^{-4}$ &  \hspace{3 pt}  $2.2 \times 10 ^{7}$ & $11.52$  \hspace{4 pt} & $2.08$  \\ 
   $6$  \hspace{5 pt} & $2.1 \times 10 ^{10}$ & $1.2 \times 10 ^{-3}$ &   \hspace{3 pt} $2.6 \times 10 ^{7}$ & $13.40$  \hspace{4 pt} & $2.26$  \\ 
   $7$  \hspace{5 pt} & $1.5 \times 10 ^{10}$ & $1.8 \times 10 ^{-3}$ &  \hspace{3 pt}  $2.8 \times 10 ^{7}$ & $14.46$  \hspace{4 pt} & $2.49$  \\ 
   $8$  \hspace{5 pt} & $1.1 \times 10 ^{10}$ & $2.4 \times 10 ^{-3}$ &   \hspace{3 pt} $2.6 \times 10 ^{7}$ & $13.81$  \hspace{4 pt} & $2.78$  \\ 
   $9$  \hspace{5 pt} & $7.4 \times 10 ^{9}$ & $3.4 \times 10 ^{-3}$  &  \hspace{3 pt}  $2.5 \times 10 ^{7}$ & $13.28$  \hspace{4 pt} & $3.19$  \\ 
 $10$  \hspace{5 pt} & $4.3 \times 10 ^{9}$ & $5.1 \times 10 ^{-3}$ &   \hspace{3 pt} $2.2 \times 10 ^{7}$ & $11.56$  \hspace{4 pt} & $3.81$  \\ 
 $11$  \hspace{5 pt} & $2.0 \times 10 ^{9}$ & $8.3 \times 10 ^{-3}$ &   \hspace{3 pt} $1.7 \times 10 ^{7}$ & $8.85$  \hspace{4 pt} & $4.89$  \\ 
 $12$  \hspace{5 pt} & $5.6 \times 10 ^{8}$ & $1.7 \times 10 ^{-2}$ &   \hspace{3 pt} $9.4 \times 10 ^{6}$ & $4.94$  \hspace{4 pt} & $7.51$  \\ 
 $13$  \hspace{5 pt} & $7.9 \times 10 ^{7}$ & $3.1 \times 10 ^{-2}$ &   \hspace{3 pt} $2.4 \times 10 ^{6}$ & $1.27$  \hspace{4 pt} & $14.47$  \\ 
 $14$ \hspace{5 pt}  & $4.0 \times 10 ^{6}$ & $5.6 \times 10 ^{-2}$ &   \hspace{3 pt} $2.3 \times 10 ^{5}$ & $0.12$  \hspace{4 pt} & $38.96$  \\ 
 $15$  \hspace{5 pt} & $4.4 \times 10 ^{4}$ & $1.0 \times 10 ^{-1}$ &   \hspace{3 pt} $4.6 \times 10 ^{3}$ & $0.00$  \hspace{4 pt} & $176.22$  \\ 
 \hline
\end{tabular}
\end{center}
\end{table}

The information combined from two means of detection measures the evolution of this density distribution as a function of time.  An imaging grid in our short flight-path apparatus, mounted perpendicular to the axis of the molecular beam, translates in $z$ to extract an electron signal waveform that traces the width of the plasma in the propagation direction as a function of its time of flight from 1 to 180 $\mu$s.  We also collect images that detail the two-dimensional projection in the $x,y$ plane, and determine the width in $z$ after nearly 500 $\mu$s of flight.  We extrapolate the slow ballistic expansion of the plasma charge distribution back to shorter evolution time of 10 $\mu$s to estimate the density distribution in a typical plasma ellipsoid, as represented by the shell model presented in Table \ref{tab:ellipse}.  Here we neglect the initial stages of bifurcation, which acts to redistribute the charge density in an ideal Gaussian ellipsoid.  A plasma with the total number of ions represented in Table \ref{tab:ellipse} retains that integrated mass for a time as long as our long flight-path instrument can measure it, at least 500 $\mu$s.  

An average of the ion density over shells determines, $\left <|{\bf r}_{ij}| \right > $, the average distance between dipoles.  $C_3^{sp}$ values computed for $\Delta n =0$ F\"oster resonant interactions in Li at $n \approx 95$ \cite{Zoubi2} provide a rough estimate of the average high-$n$ dipole-dipole matrix element, $t$.  Combining these quantities yields an upper-limiting estimate of $\widetilde{J}$.

We can expect the perturbation of individual molecules by the background of charged-particles in the plasma.  This will diminish the probability of finding resonant target states and decrease the value of $\widetilde{J}$ for symmetric exchange.  Thus, we can expect a quenched system in which dipole-dipole interactions occur with rarity and randomness, sampling the huge state space defined by the measured distribution of electron binding energies, $W$.  

As noted above in the caption to Figure \ref{fig:n44SFI}, the SFI spectrum, which displays a distribution of electron binding energies in the arrested state of the plasma, provides a direct measure of $W$.  Table \ref{tab:summary} summarizes this information together with other properties of the quenched plasma derived from experiment, including $\widetilde{J}$, as estimated for Li under our conditions as an upper limit.

\begin{table}[h!]
 \caption{Resonance counting parameters under conditions of arrested relaxation in the quenched ultracold plasma.  We take the disorder width $W$ directly from the width of the plasma feature in the SIF spectrum.  $\widetilde{J}$ is derived as a rough upper-limiting estimate, based on the average dipole-dipole matrix element, $t$ computed for $\Delta n =0$ interactions in alkali metals \cite{Zoubi2}, together with the mean distance between NO$^+$ ions in the shell model ellipsoid.  These parameters determine $N_c$, a critical number of dipoles required for delocalization.  $R_c$ describes the length scale for delocalization and $\tau_c$ denotes the delocalization time, given a sufficient number of dipoles at the average density of the experiment.  Note that the ultracold plasma quenched experimentally contains an order of magnitude fewer than $N_c$ dipoles.  } 
  \label{tab:summary}
  \begin{center}
 \begin{tabular}{ccccccc}  
 \toprule 
$W$   & $ t$ &    \hspace{1 pt}   $\left <|{\bf r}_{ij}| \right > $     \hspace{1 pt}  &   $\widetilde{J}$  &   $N_c$   &  $R_c$   &     $\tau_c$  \\ 
GHz      & \hspace{1 pt} GHz($\mu$m)$^3$   \hspace{1 pt}  &   $\mu$m   &    GHz     &       &   $\mu$m     &    s   \\
 \hline
 500  &  75  &  3.3  &  2.0  &  \hspace{3 pt} $3.6 \times 10^9$  \hspace{3 pt} &   4000  &  0.85  \\
 \hline 
\end{tabular}
\end{center}
\end{table}

In the locator expansion picture, a system with arbitrary disorder delocalizes whenever the number of dipoles exceeds a critical number, $N_c$, determined by the disorder width, $W$ and the average coupling strength, $\widetilde J$.  The coupling terms in the Hamiltonian defined by Eq (\ref{XY-Ising}) scale with ${R}$ in a three-dimensional system to set a critical number of dipoles defined by a quantity, $N_c = (W/\widetilde{J})^4$ derived above.  Thus, a quenched state described by the elliptical shell model, with a density distribution as in Table \ref{tab:ellipse}, a measured $W = 500$ GHz, and an upper limiting estimated $\widetilde{J}= 2$ GHz, requires $N_c = 3.6 \times 10^9$ interacting dipoles to delocalize.  This number of dipoles, deemed necessary for resonance delocalization, exceeds the experimental number of excited molecules in the quenched ultracold plasma by more than an order of magnitude.

Returning to Eq (\ref{Nc4}), we can figure for a system with a peak density of dipoles, $\rho = 4 \times 10^{10}~{\rm cm}^{-3}$ that a threshold a critical number of dipoles, $N_c = 3.6 \times 10^9$ must attain a critical interaction distance or minimum system size, $R_c \approx 4$ mm, to delocalize.  For such a range, an upper-limiting dipole-dipole matrix element, $ \left < t_{ij} \right > = 75$ GHz($\mu$m)$^3$, calls for an irreversible transition time, $\tau_c$, of about 1 second \cite{Burin2}.  Note that the quenched ultracold plasma volume formed experimentally contains an order of magnitude fewer dipoles than $N_c$, as determined by the model of reference \cite{Burin1}.  

While the experimental plasma detailed above forms with a well-defined total number of dipoles, we cannot describe its constituent quantum states or their dipole-dipole interaction with precision.  This limits the accuracy with which we can specify a value of $N_c$. Moreover, it remains to be seen whether resonance-counting arguments can serve to define the limiting conditions for MBL in higher dimensions \cite{Sondhi}.  

{\color{tempcolor}{Inevitable expansion of the plasma modifies this picture, as higher-$L$ interactions accumulate and effects of motion possibly change the physics.  Nevertheless, we expect that the strongly quenched state of localization in the initial-time evolution to lead to, at least, very slow dynamics on the observational timescales.}}

\subsection{Beyond locator-expansion approaches}

The analysis above follows from a simple perturbative argument in which a diverging number of resonances calls for a breakdown of the locator expansion, signalling delocalization. However, Nandkishore and Sondhi have pointed out \cite{Sondhi} that a break-down of the locator expansion need not imply an absence of localization.  In their work, they studied a long-range Coulomb model for which they argue that non-perturbative short-range interactions can arise in a regime for which locator expansion arguments can be correctly applied, concluding that many-body localization may well extend to long-range interacting systems of higher dimension \cite{Safavi}.

We have presented a conservative model based on resonance counting arguments that argues for disorder-driven localization in absence of off-diagonal disorder (in $J_{ij}$).  The disorder in $J_{ij}$ ought to lead to unconventional resonance structure, that will very likely prevent delocalization, as it becomes very difficult for excitations to propagate amongst a chain of dipoles involving very different quantum states and very different transition amplitudes. Moreover, extensions studying non-perturbative effects in the spirit of Nandkishore and Sondhi, while not easily generalizable to dipolar systems, seem to suggest that if localization survives Coulomb interactions, it should in principle survive shorter-range dipolar interactions. 

\subsection{Rare thermal regions}
Our analysis completely ignores the potential effects of so-called Griffiths region. These rare regions are thought to appear spontaneously as a product of delocalizing fluctuations in the localized phase, forming thermal reservoirs that ultimately destabilize MBL systems of higher dimension \cite{Sarang1, DeRoeck,Roeck_griffith, RareRegions_rev, Thermal_inclusions,potirniche2018stability}. These regions lead to a delocalization scenario in which the thermal region grows, driving the MBL phase to form a glass, that evolves at best with slow dynamics.  

An important element of the self-assembly of the molecular ultracold plasma may diminish the effect of rare thermal regions as a destabilizing factor.  As the quenched ultracold plasma approaches a state of arrested relaxation, SFI spectra show that electron-Rydberg collisions cease to drive energy transport, defying classical models of plasma kinetics. If a localized plasma happens to develop a Griffiths region, thermal relaxation can be expected to produce a local volume of electron gas, heated by inelastic collisions with Rydberg molecules.  These collisions would rapidly transport the Rydberg population to a domain of low $n$, where strong intramolecular Rydberg-valence coupling causes rapid predissociation \cite{Giusti,Remacle}.  Such a loss of Rydberg NO to form neutral N($^4$S) and O($^3$P) would irreversibly dissipate the thermal energy of the delocalized region.  

\section{Conclusions}

Experiment paints a very clear picture of arrested relaxation in an ultracold neutral plasma of NO$^+$ ions and electrons.  Formed by electron-impact avalanche in a state-selected Rydberg gas of supersonically cooled nitric oxide, this plasma spontaneously bifurcates, quenching the energy distributions of ions and electrons to form compressed volumes of canonical density and ultracold temperature.  Experimental images detail the time evolution of the electron binding energy and three-dimensional density distribution.  These metrics directly show evidence for an initial redistribution of Rydberg population driven by collision with energetic electrons.  Thereafter, all sign of free electrons disappears, and the relaxation of the system ceases.  These dynamics fail to conform with conventional models of plasma coupled-rate processes and NO predissociation.  

The present work proceeds from this well-defined base of experimental observations to formulate a conceptual rationale for localization as it might occur in an ensemble of strongly interacting random dipoles.  We refer to the selective field ionization (SFI) spectrum of electron binding energies to gauge the ensemble disorder in the energies associated with dipole-dipole resonant couplings, and construct a scheme to describe the resulting web of interactions in terms a projection to some low number, $L$, of random coupled transitions.  The vast array of accessible pathways, and the randomness of the system of states populated in the quenched plasma far exceeds the complexity typically encountered in few-level quantum optics systems.  Nevertheless, we take the simplest possible approach of an XY Hamiltonian for interacting dipolar spins with emergent Ising interactions.  Applying the idea of a locator expansion, we consider the conditions required for delocalization in an $L=2$ limit.  We find $N_c$, a critical number of dipoles required to sustain delocalization, that exceeds the number of dipoles in the plasma by more than an order of magnitude.  We thus argue that the plasma ought to exhibit localization on a long observational timescale at least.  

We represent this quenched ultracold plasma as a class of system in which localization arises naturally, in which locally emergent conservation laws act in concert with a quench to guide the propagation of the system toward a global many-body localized state.  This evolution must draw upon a subtle interplay between dissociation and resonant but rare dipole-dipole interactions giving rise to novel many-body phenomena.  Tracking the state of the molecular ensemble, as it proceeds from the semi-classical conditions of a spectroscopically defined Rydberg gas, provides a unique opportunity to observe stages in the development of localization.  While the construction of a picture of the full many-body Fock space of the random Rydberg/excitonic dipoles will require new theoretical tools, strong evidence of dipole-dipole interactions remains, providing a logical basis for the projection to a simple leading order of resonance and a minimal effective model for localization.  Further work is required to wrestle with the theoretical complexity of the quenched plasma.  Effects such as Griffiths regions and dissociation may prove to have singular importance in this molecular system.  The emergence of discerning experimental tools promise both greater fundamental understanding and perhaps important new applications.

\section*{Acknowledgments}
We gratefully acknowledge useful discussions with Rahul Nandkishore, Shivaji Sondhi, Ehud Altman, Sid Parmeswaran, Steve Simon, Antonello Scardicchio, Romain Vasseur and Hossein Sadeghpour. This work was supported by the US Air Force Office of Scientific Research (Grant No. FA9550-17-1-0343), together with the Natural Sciences and Engineering research Council of Canada (NSERC), the Canada Foundation for Innovation (CFI) and the British Columbia Knowledge Development Fund (BCKDF).

\section*{References}


\newcommand{\BMCxmlcomment}[1]{}

\BMCxmlcomment{

<refgrp>

<bibl id="B1">
  <title><p>{Quantum beat spectroscopy in chemistry}</p></title>
  <aug>
    <au><snm>Carter</snm><fnm>RT</fnm></au>
    <au><snm>Huber</snm><fnm>JR</fnm></au>
  </aug>
  <source>{Chem Soc Rev}</source>
  <pubdate>{2000}</pubdate>
  <volume>{29}</volume>
  <issue>{5}</issue>
  <fpage>{305</fpage>
  <lpage>314}</lpage>
</bibl>

<bibl id="B2">
  <title><p>{Quantum ergodicity and energy flow in molecules}</p></title>
  <aug>
    <au><snm>Leitner</snm><fnm>DM</fnm></au>
  </aug>
  <source>Adv Phys</source>
  <pubdate>{2015}</pubdate>
  <volume>{64}</volume>
  <issue>{4}</issue>
  <fpage>{445</fpage>
  <lpage>517}</lpage>
</bibl>

<bibl id="B3">
  <title><p>Quantum statistical mechanics in a closed system</p></title>
  <aug>
    <au><snm>Deutsch</snm><fnm>J. M.</fnm></au>
  </aug>
  <source>Phys. Rev. A</source>
  <publisher>American Physical Society</publisher>
  <pubdate>1991</pubdate>
  <volume>43</volume>
  <fpage>2046</fpage>
  <lpage>-2049</lpage>
</bibl>

<bibl id="B4">
  <title><p>Chaos and quantum thermalization</p></title>
  <aug>
    <au><snm>Srednicki</snm><fnm>M</fnm></au>
  </aug>
  <source>Phys. Rev. E</source>
  <publisher>American Physical Society</publisher>
  <pubdate>1994</pubdate>
  <volume>50</volume>
  <fpage>888</fpage>
  <lpage>-901</lpage>
</bibl>

<bibl id="B5">
  <title><p>From Quantum Dynamics to the Canonical Distribution: General
  Picture and a Rigorous Example</p></title>
  <aug>
    <au><snm>Tasaki</snm><fnm>H</fnm></au>
  </aug>
  <source>Phys. Rev. Lett.</source>
  <publisher>American Physical Society</publisher>
  <pubdate>1998</pubdate>
  <volume>80</volume>
  <fpage>1373</fpage>
  <lpage>-1376</lpage>
</bibl>

<bibl id="B6">
  <title><p>Thermalization and its mechanism for generic isolated quantum
  systems</p></title>
  <aug>
    <au><snm>Rigol</snm><fnm>M</fnm></au>
    <au><snm>Dunjko</snm><fnm>V</fnm></au>
    <au><snm>Olshanii</snm><fnm>M</fnm></au>
  </aug>
  <source>Nature</source>
  <pubdate>2008</pubdate>
  <volume>452</volume>
  <fpage>854</fpage>
  <lpage>-858</lpage>
</bibl>

<bibl id="B7">
  <title><p>Quantum many-body systems out of equilibrium</p></title>
  <aug>
    <au><snm>Eisert</snm><fnm>J.</fnm></au>
    <au><snm>Friesdorf</snm><fnm>M.</fnm></au>
    <au><snm>Gogolin</snm><fnm>C.</fnm></au>
  </aug>
  <source>Nature Physics</source>
  <pubdate>2015</pubdate>
  <volume>11</volume>
  <fpage>124</fpage>
  <lpage>-130</lpage>
</bibl>

<bibl id="B8">
  <title><p>From quantum chaos and eigenstate thermalization to statistical
  mechanics and thermodynamics</p></title>
  <aug>
    <au><snm>D'Alessio</snm><fnm>L</fnm></au>
    <au><snm>Kafri</snm><fnm>Y</fnm></au>
    <au><snm>Polkovnikov</snm><fnm>A</fnm></au>
    <au><snm>Rigol</snm><fnm>M</fnm></au>
  </aug>
  <source>Adv. Phys.</source>
  <pubdate>2016</pubdate>
  <volume>65</volume>
  <issue>3</issue>
  <fpage>239</fpage>
  <lpage>362</lpage>
</bibl>

<bibl id="B9">
  <title><p>Ergodic dynamics and thermalization in an isolated quantum
  system</p></title>
  <aug>
    <au><snm>Neill</snm><fnm>C.</fnm></au>
    <au><snm>Roushan</snm><fnm>P.</fnm></au>
    <au><snm>Fang</snm><fnm>M.</fnm></au>
    <au><snm>Chen</snm><fnm>Y.</fnm></au>
    <au><snm>Kolodrubetz</snm><fnm>M.</fnm></au>
    <au><snm>Chen</snm><fnm>Z.</fnm></au>
    <au><snm>Megrant</snm><fnm>A.</fnm></au>
    <au><snm>Barends</snm><fnm>R.</fnm></au>
    <au><snm>Campbell</snm><fnm>B.</fnm></au>
    <au><snm>Chiaro</snm><fnm>B.</fnm></au>
    <au><snm>Dunsworth</snm><fnm>A.</fnm></au>
    <au><snm>Jeffrey</snm><fnm>E.</fnm></au>
    <au><snm>Kelly</snm><fnm>J.</fnm></au>
    <au><snm>Mutus</snm><fnm>J.</fnm></au>
    <au><snm>O'Malley</snm><fnm>P. J. J.</fnm></au>
    <au><snm>Quintana</snm><fnm>C.</fnm></au>
    <au><snm>Sank</snm><fnm>D.</fnm></au>
    <au><snm>Vainsencher</snm><fnm>A.</fnm></au>
    <au><snm>Wenner</snm><fnm>J.</fnm></au>
    <au><snm>White</snm><fnm>T. C.</fnm></au>
    <au><snm>Polkovnikov</snm><fnm>A.</fnm></au>
    <au><snm>Martinis</snm><fnm>J. M.</fnm></au>
  </aug>
  <source>Nature Physics</source>
  <pubdate>2016</pubdate>
  <volume>12</volume>
  <fpage>1037</fpage>
  <lpage>-1041</lpage>
</bibl>

<bibl id="B10">
  <title><p>Unimolecular Reaction Dynamics</p></title>
  <aug>
    <au><snm>Baer</snm><fnm>T</fnm></au>
    <au><snm>Hase</snm><fnm>W L</fnm></au>
  </aug>
  <publisher>New York: Oxford University Press</publisher>
  <pubdate>1996</pubdate>
</bibl>

<bibl id="B11">
  <title><p>{Interacting Electrons in Disordered Wires: Anderson Localization
  and Low-$T$ Transport}</p></title>
  <aug>
    <au><snm>Gornyi</snm><fnm>I. V.</fnm></au>
    <au><snm>Mirlin</snm><fnm>A. D.</fnm></au>
    <au><snm>Polyakov</snm><fnm>D. G.</fnm></au>
  </aug>
  <source>Phys Rev Lett</source>
  <pubdate>2005</pubdate>
  <volume>95</volume>
  <fpage>206603</fpage>
</bibl>

<bibl id="B12">
  <title><p>Metal--insulator transition in a weakly interacting many-electron
  system with localized single-particle states</p></title>
  <aug>
    <au><snm>Basko</snm><fnm>D.M.</fnm></au>
    <au><snm>Aleiner</snm><fnm>I.L.</fnm></au>
    <au><snm>Altshuler</snm><fnm>B.L.</fnm></au>
  </aug>
  <source>Ann. Phys.</source>
  <pubdate>2006</pubdate>
  <volume>321</volume>
  <issue>5</issue>
  <fpage>1126</fpage>
  <lpage>1205</lpage>
</bibl>

<bibl id="B13">
  <title><p>Many-body localization and thermalization in quantum statistical
  mechanics</p></title>
  <aug>
    <au><snm>Nandkishore</snm><fnm>R</fnm></au>
    <au><snm>Huse</snm><fnm>DA</fnm></au>
  </aug>
  <source>Annu. Rev. Condens. Matter Phys.</source>
  <publisher>Annual Reviews</publisher>
  <pubdate>2015</pubdate>
  <volume>6</volume>
  <issue>1</issue>
  <fpage>15</fpage>
  <lpage>-38</lpage>
</bibl>

<bibl id="B14">
  <title><p>Many-body localization: An introduction and selected
  topics</p></title>
  <aug>
    <au><snm>Alet</snm><fnm>F</fnm></au>
    <au><snm>Laflorencie</snm><fnm>N</fnm></au>
  </aug>
  <source>Comptes Rendus Physique</source>
  <pubdate>2018</pubdate>
</bibl>

<bibl id="B15">
  <title><p>Many-body localization, symmetry and topology</p></title>
  <aug>
    <au><snm>Parameswaran</snm><fnm>S A</fnm></au>
    <au><snm>Vasseur</snm><fnm>R</fnm></au>
  </aug>
  <source>Reports on Progress in Physics</source>
  <pubdate>2018</pubdate>
  <volume>81</volume>
  <issue>8</issue>
  <fpage>082501</fpage>
</bibl>

<bibl id="B16">
  <title><p>Ergodicity, Entanglement and Many-Body Localization</p></title>
  <aug>
    <au><snm>Abanin</snm><fnm>DA</fnm></au>
    <au><snm>Altman</snm><fnm>E</fnm></au>
    <au><snm>Bloch</snm><fnm>I</fnm></au>
    <au><snm>Serbyn</snm><fnm>M</fnm></au>
  </aug>
  <source>arXiv preprint arXiv:1804.11065</source>
  <pubdate>2018</pubdate>
</bibl>

<bibl id="B17">
  <title><p>Many-body localization protected quantum state transfer</p></title>
  <aug>
    <au><snm>Yao</snm><fnm>NY</fnm></au>
    <au><snm>Laumann</snm><fnm>CR</fnm></au>
    <au><snm>Vishwanath</snm><fnm>A</fnm></au>
  </aug>
  <source>arXiv preprint arXiv:1508.06995</source>
  <pubdate>2015</pubdate>
</bibl>

<bibl id="B18">
  <title><p>Dynamics of quantum information in many-body localized
  systems</p></title>
  <aug>
    <au><snm>Ba\ nuls</snm><fnm>M. C.</fnm></au>
    <au><snm>Yao</snm><fnm>N. Y.</fnm></au>
    <au><snm>Choi</snm><fnm>S.</fnm></au>
    <au><snm>Lukin</snm><fnm>M. D.</fnm></au>
    <au><snm>Cirac</snm><fnm>J. I.</fnm></au>
  </aug>
  <source>Phys. Rev. B</source>
  <pubdate>2017</pubdate>
  <volume>96</volume>
  <fpage>174201</fpage>
</bibl>

<bibl id="B19">
  <title><p>Disorder-Induced Localization in a Strongly Correlated Atomic
  Hubbard Gas</p></title>
  <aug>
    <au><snm>Kondov</snm><fnm>S. S.</fnm></au>
    <au><snm>McGehee</snm><fnm>W. R.</fnm></au>
    <au><snm>Xu</snm><fnm>W.</fnm></au>
    <au><snm>DeMarco</snm><fnm>B.</fnm></au>
  </aug>
  <source>Phys. Rev. Lett.</source>
  <publisher>American Physical Society</publisher>
  <pubdate>2015</pubdate>
  <volume>114</volume>
  <fpage>083002</fpage>
</bibl>

<bibl id="B20">
  <title><p>{Observation of many-body localization of interacting fermions in a
  quasi-random optical lattice.}</p></title>
  <aug>
    <au><snm>Schreiber</snm><fnm>M</fnm></au>
    <au><snm>Hodgman</snm><fnm>S S</fnm></au>
    <au><snm>Bordia</snm><fnm>P</fnm></au>
    <au><snm>L{\"u}schen</snm><fnm>H P</fnm></au>
    <au><snm>Fischer</snm><fnm>M H</fnm></au>
    <au><snm>Vosk</snm><fnm>R</fnm></au>
    <au><snm>Altman</snm><fnm>E</fnm></au>
    <au><snm>Schneider</snm><fnm>U</fnm></au>
    <au><snm>Bloch</snm><fnm>I</fnm></au>
  </aug>
  <source>Science</source>
  <pubdate>2015</pubdate>
  <volume>349</volume>
  <fpage>842</fpage>
  <lpage>845</lpage>
</bibl>

<bibl id="B21">
  <title><p>Exploring the many-body localization transition in two
  dimensions</p></title>
  <aug>
    <au><snm>Choi</snm><fnm>JY</fnm></au>
    <au><snm>Hild</snm><fnm>S</fnm></au>
    <au><snm>Zeiher</snm><fnm>J</fnm></au>
    <au><snm>Schau{\ss}</snm><fnm>P</fnm></au>
    <au><snm>Rubio Abadal</snm><fnm>A</fnm></au>
    <au><snm>Yefsah</snm><fnm>T</fnm></au>
    <au><snm>Khemani</snm><fnm>V</fnm></au>
    <au><snm>Huse</snm><fnm>DA</fnm></au>
    <au><snm>Bloch</snm><fnm>I</fnm></au>
    <au><snm>Gross</snm><fnm>C</fnm></au>
  </aug>
  <source>Science</source>
  <pubdate>2016</pubdate>
  <volume>352</volume>
  <issue>6293</issue>
  <fpage>1547</fpage>
  <lpage>-1552</lpage>
</bibl>

<bibl id="B22">
  <title><p>Probing Slow Relaxation and Many-Body Localization in
  Two-Dimensional Quasi-Periodic Systems</p></title>
  <aug>
    <au><snm>Bordia</snm><fnm>P</fnm></au>
    <au><snm>L{\"u}schen</snm><fnm>H</fnm></au>
    <au><snm>Scherg</snm><fnm>S</fnm></au>
    <au><snm>Gopalakrishnan</snm><fnm>S</fnm></au>
    <au><snm>Knap</snm><fnm>M</fnm></au>
    <au><snm>Schneider</snm><fnm>U</fnm></au>
    <au><snm>Bloch</snm><fnm>I</fnm></au>
  </aug>
  <source>Phys. Rev. X</source>
  <pubdate>2017</pubdate>
  <volume>7</volume>
  <fpage>041047</fpage>
</bibl>

<bibl id="B23">
  <title><p>Observation of Slow Dynamics near the Many-Body Localization
  Transition in One-Dimensional Quasiperiodic Systems</p></title>
  <aug>
    <au><snm>L\"uschen</snm><fnm>HP</fnm></au>
    <au><snm>Bordia</snm><fnm>P</fnm></au>
    <au><snm>Scherg</snm><fnm>S</fnm></au>
    <au><snm>Alet</snm><fnm>F</fnm></au>
    <au><snm>Altman</snm><fnm>E</fnm></au>
    <au><snm>Schneider</snm><fnm>U</fnm></au>
    <au><snm>Bloch</snm><fnm>I</fnm></au>
  </aug>
  <source>Phys. Rev. Lett.</source>
  <pubdate>2017</pubdate>
  <volume>119</volume>
  <fpage>260401</fpage>
</bibl>

<bibl id="B24">
  <title><p>Signatures of Many-Body Localization in a Controlled Open Quantum
  System</p></title>
  <aug>
    <au><snm>L\"uschen</snm><fnm>HP</fnm></au>
    <au><snm>Bordia</snm><fnm>P</fnm></au>
    <au><snm>Hodgman</snm><fnm>SS</fnm></au>
    <au><snm>Schreiber</snm><fnm>M</fnm></au>
    <au><snm>Sarkar</snm><fnm>S</fnm></au>
    <au><snm>Daley</snm><fnm>AJ</fnm></au>
    <au><snm>Fischer</snm><fnm>MH</fnm></au>
    <au><snm>Altman</snm><fnm>E</fnm></au>
    <au><snm>Bloch</snm><fnm>I</fnm></au>
    <au><snm>Schneider</snm><fnm>U</fnm></au>
  </aug>
  <source>Phys. Rev. X</source>
  <publisher>American Physical Society</publisher>
  <pubdate>2017</pubdate>
  <volume>7</volume>
  <fpage>011034</fpage>
</bibl>

<bibl id="B25">
  <title><p>Many-body localization in a quantum simulator with programmable
  random disorder</p></title>
  <aug>
    <au><snm>Smith</snm><fnm>J.</fnm></au>
    <au><snm>Lee</snm><fnm>A.</fnm></au>
    <au><snm>Richerme</snm><fnm>P.</fnm></au>
    <au><snm>Neyenhuis</snm><fnm>B.</fnm></au>
    <au><snm>Hess</snm><fnm>P. W.</fnm></au>
    <au><snm>Hauke</snm><fnm>P.</fnm></au>
    <au><snm>Heyl</snm><fnm>M.</fnm></au>
    <au><snm>Huse</snm><fnm>D. A.</fnm></au>
    <au><snm>Monroe</snm><fnm>C.</fnm></au>
  </aug>
  <source>Nat. Phys.</source>
  <pubdate>2016</pubdate>
  <volume>12</volume>
  <issue>10</issue>
  <fpage>907</fpage>
  <lpage>-911</lpage>
</bibl>

<bibl id="B26">
  <title><p>Probing many-body localization in the presence of a quantum
  bath</p></title>
  <aug>
    <au><snm>Rubio Abadal</snm><fnm>A</fnm></au>
    <au><snm>Choi</snm><fnm>Jy</fnm></au>
    <au><snm>Zeiher</snm><fnm>J</fnm></au>
    <au><snm>Hollerith</snm><fnm>S</fnm></au>
    <au><snm>Rui</snm><fnm>J</fnm></au>
    <au><snm>Bloch</snm><fnm>I</fnm></au>
    <au><snm>Gross</snm><fnm>C</fnm></au>
  </aug>
  <source>arXiv preprint arXiv:1805.00056</source>
  <pubdate>2018</pubdate>
</bibl>

<bibl id="B27">
  <title><p>Probing entanglement in a many-body-localized system</p></title>
  <aug>
    <au><snm>Lukin</snm><fnm>A</fnm></au>
    <au><snm>Rispoli</snm><fnm>M</fnm></au>
    <au><snm>Schittko</snm><fnm>R</fnm></au>
    <au><snm>Tai</snm><fnm>ME</fnm></au>
    <au><snm>Kaufman</snm><fnm>AM</fnm></au>
    <au><snm>Choi</snm><fnm>S</fnm></au>
    <au><snm>Khemani</snm><fnm>V</fnm></au>
    <au><snm>L{\'e}onard</snm><fnm>J</fnm></au>
    <au><snm>Greiner</snm><fnm>M</fnm></au>
  </aug>
  <source>arXiv preprint arXiv:1805.09819</source>
  <pubdate>2018</pubdate>
</bibl>

<bibl id="B28">
  <title><p>Local Conservation Laws and the Structure of the Many-Body
  Localized States</p></title>
  <aug>
    <au><snm>Serbyn</snm><fnm>M</fnm></au>
    <au><snm>\'{c}\fi{}</snm><fnm>Z.</fnm></au>
    <au><snm>Abanin</snm><fnm>DA</fnm></au>
  </aug>
  <source>Phys. Rev. Lett.</source>
  <publisher>American Physical Society</publisher>
  <pubdate>2013</pubdate>
  <volume>111</volume>
  <fpage>127201</fpage>
</bibl>

<bibl id="B29">
  <title><p>Phenomenology of fully many-body-localized systems</p></title>
  <aug>
    <au><snm>Huse</snm><fnm>DA</fnm></au>
    <au><snm>Nandkishore</snm><fnm>R</fnm></au>
    <au><snm>Oganesyan</snm><fnm>V</fnm></au>
  </aug>
  <source>Phys. Rev. B</source>
  <publisher>American Physical Society</publisher>
  <pubdate>2014</pubdate>
  <volume>90</volume>
  <fpage>174202</fpage>
</bibl>

<bibl id="B30">
  <title><p>Constructing local integrals of motion in the many-body localized
  phase</p></title>
  <aug>
    <au><snm>Chandran</snm><fnm>A</fnm></au>
    <au><snm>Kim</snm><fnm>IH</fnm></au>
    <au><snm>Vidal</snm><fnm>G</fnm></au>
    <au><snm>Abanin</snm><fnm>DA</fnm></au>
  </aug>
  <source>Phys. Rev. B</source>
  <publisher>American Physical Society</publisher>
  <pubdate>2015</pubdate>
  <volume>91</volume>
  <fpage>085425</fpage>
</bibl>

<bibl id="B31">
  <title><p>Integrals of motion in the many-body localized phase</p></title>
  <aug>
    <au><snm>Ros</snm><fnm>V.</fnm></au>
    <au><snm>M{\"u}ller</snm><fnm>M.</fnm></au>
    <au><snm>Scardicchio</snm><fnm>A.</fnm></au>
  </aug>
  <source>Nuclear Physics B</source>
  <pubdate>2015</pubdate>
  <volume>891</volume>
  <fpage>420</fpage>
  <lpage>465</lpage>
</bibl>

<bibl id="B32">
  <title><p>Local integrals of motion in many-body localized
  systems</p></title>
  <aug>
    <au><snm>Imbrie</snm><fnm>JZ</fnm></au>
    <au><snm>Ros</snm><fnm>V</fnm></au>
    <au><snm>Scardicchio</snm><fnm>A</fnm></au>
  </aug>
  <source>Annalen der Physik</source>
  <pubdate>2017</pubdate>
  <volume>529</volume>
  <issue>7</issue>
  <fpage>1600278</fpage>
</bibl>

<bibl id="B33">
  <title><p>Evolution from a molecular Rydberg gas to an ultracold plasma in a
  seeded supersonic expansion of NO</p></title>
  <aug>
    <au><snm>Morrison</snm><fnm>J P</fnm></au>
    <au><snm>Rennick</snm><fnm>C J</fnm></au>
    <au><snm>Keller</snm><fnm>J S</fnm></au>
    <au><snm>Grant</snm><fnm>E R</fnm></au>
  </aug>
  <source>Phys. Rev. Lett.</source>
  <pubdate>2008</pubdate>
  <volume>101</volume>
  <fpage>205005</fpage>
</bibl>

<bibl id="B34">
  <title><p>{Molecular ion--electron recombination in an expanding ultracold
  neutral plasma of NO$^+$}</p></title>
  <aug>
    <au><snm>Sadeghi</snm><fnm>H</fnm></au>
    <au><snm>Schulz Weiling</snm><fnm>M</fnm></au>
    <au><snm>Morrison</snm><fnm>JP</fnm></au>
    <au><snm>Yiu</snm><fnm>JCH</fnm></au>
    <au><snm>Saquet</snm><fnm>N</fnm></au>
    <au><snm>Rennick</snm><fnm>CJ</fnm></au>
    <au><snm>Grant</snm><fnm>E</fnm></au>
  </aug>
  <source>Phys Chem Chem Phys</source>
  <pubdate>2011</pubdate>
  <volume>13</volume>
  <fpage>18872</fpage>
</bibl>

<bibl id="B35">
  <title><p>{Recombinative dissociation and the evolution of a molecular
  ultracold plasma}</p></title>
  <aug>
    <au><snm>Saquet</snm><fnm>N</fnm></au>
    <au><snm>Morrison</snm><fnm>JP</fnm></au>
    <au><snm>Grant</snm><fnm>E</fnm></au>
  </aug>
  <source>J Phys B</source>
  <pubdate>2012</pubdate>
  <volume>45</volume>
  <fpage>175302</fpage>
</bibl>

<bibl id="B36">
  <title><p>Dissociation and the development of spatial correlation in a
  molecular ultracold plasma</p></title>
  <aug>
    <au><snm>Sadeghi</snm><fnm>H</fnm></au>
    <au><snm>Kruyen</snm><fnm>A</fnm></au>
    <au><snm>Hung</snm><fnm>J</fnm></au>
    <au><snm>Gurian</snm><fnm>J H</fnm></au>
    <au><snm>Morrison</snm><fnm>J P</fnm></au>
    <au><snm>Schulz Weiling</snm><fnm>M</fnm></au>
    <au><snm>Saquet</snm><fnm>N</fnm></au>
    <au><snm>Rennick</snm><fnm>C J</fnm></au>
    <au><snm>Grant</snm><fnm>E R</fnm></au>
  </aug>
  <source>Phys Rev Lett</source>
  <pubdate>2014</pubdate>
  <volume>112</volume>
  <fpage>075001</fpage>
</bibl>

<bibl id="B37">
  <title><p>Quantum state control of ultracold plasma fission</p></title>
  <aug>
    <au><snm>Schulz Weiling</snm><fnm>M</fnm></au>
    <au><snm>Grant</snm><fnm>E R</fnm></au>
  </aug>
  <source>J Phys B</source>
  <pubdate>2016</pubdate>
  <volume>49</volume>
  <fpage>064009</fpage>
</bibl>

<bibl id="B38">
  <title><p>Arrested relaxation in an isolated molecular ultracold
  plasma</p></title>
  <aug>
    <au><snm>Haenel</snm><fnm>R</fnm></au>
    <au><snm>Schulz Weiling</snm><fnm>M</fnm></au>
    <au><snm>Sous</snm><fnm>J</fnm></au>
    <au><snm>Sadeghi</snm><fnm>H</fnm></au>
    <au><snm>Aghigh</snm><fnm>M</fnm></au>
    <au><snm>Melo</snm><fnm>L</fnm></au>
    <au><snm>Keller</snm><fnm>JS</fnm></au>
    <au><snm>Grant</snm><fnm>ER</fnm></au>
  </aug>
  <source>Phys Rev A</source>
  <publisher>APS</publisher>
  <pubdate>2017</pubdate>
  <volume>96</volume>
  <fpage>023613</fpage>
</bibl>

<bibl id="B39">
  <title><p>On the formation and decay of a molecular ultracold
  plasma</p></title>
  <aug>
    <au><snm>Saquet</snm><fnm>N.</fnm></au>
    <au><snm>Morrison</snm><fnm>J. P.</fnm></au>
    <au><snm>Schulz Weiling</snm><fnm>M.</fnm></au>
    <au><snm>Sadeghi</snm><fnm>H.</fnm></au>
    <au><snm>Yiu</snm><fnm>J.</fnm></au>
    <au><snm>Rennick</snm><fnm>C. J.</fnm></au>
    <au><snm>Grant</snm><fnm>E. R.</fnm></au>
  </aug>
  <source>J Phys B</source>
  <pubdate>2011</pubdate>
  <volume>44</volume>
  <fpage>184015</fpage>
</bibl>

<bibl id="B40">
  <title><p>Coupled rate-equation hydrodynamic simulation of a Rydberg gas
  Gaussian ellipsoid: Classical avalanche and evolution to molecular
  plasma</p></title>
  <aug>
    <au><snm>Haenel</snm><fnm>R</fnm></au>
    <au><snm>Grant</snm><fnm>E R</fnm></au>
  </aug>
  <source>Chem Phys</source>
  <pubdate>2018</pubdate>
  <volume>514</volume>
  <fpage>55</fpage>
  <lpage>66</lpage>
</bibl>

<bibl id="B41">
  <title><p>Possible Many-Body Localization in a Long-Lived Finite-Temperature
  Ultracold Quasineutral Molecular Plasma</p></title>
  <aug>
    <au><snm>Sous</snm><fnm>J</fnm></au>
    <au><snm>Grant</snm><fnm>E</fnm></au>
  </aug>
  <source>Phys. Rev. Lett.</source>
  <publisher>American Physical Society</publisher>
  <pubdate>2018</pubdate>
  <volume>120</volume>
  <fpage>110601</fpage>
</bibl>

<bibl id="B42">
  <title><p>{Kinetic modeling and molecular dynamics simulation of ultracold
  neutral plasmas including ionic correlations}</p></title>
  <aug>
    <au><snm>Pohl</snm><fnm>T</fnm></au>
    <au><snm>Pattard</snm><fnm>T</fnm></au>
    <au><snm>Rost</snm><fnm>J M</fnm></au>
  </aug>
  <source>Physical Review A</source>
  <pubdate>2004</pubdate>
  <volume>70</volume>
  <issue>3</issue>
  <fpage>33416</fpage>
</bibl>

<bibl id="B43">
  <title><p>{Molecular Applications of Quantum Defect Theory}</p></title>
  <aug>
    <au><snm>Greene</snm><fnm>CH</fnm></au>
    <au><snm>Jungen</snm><fnm>C</fnm></au>
  </aug>
  <source>Adv in Atomic and Molecular Phys Volume 21</source>
  <publisher>Elsevier</publisher>
  <pubdate>1985</pubdate>
  <fpage>51</fpage>
  <lpage>-121</lpage>
</bibl>

<bibl id="B44">
  <title><p>{Theoretical study of competing photoionization and
  photodissociation processes in the NO molecule}</p></title>
  <aug>
    <au><snm>Giusti Suzor</snm><fnm>A</fnm></au>
    <au><snm>Jungen</snm><fnm>C</fnm></au>
  </aug>
  <source>J Chem Phys</source>
  <pubdate>1984</pubdate>
  <volume>80</volume>
  <issue>3</issue>
  <fpage>986</fpage>
</bibl>

<bibl id="B45">
  <title><p>{Multipole expansion in plasmas: Effective interaction potentials
  between compound particles}</p></title>
  <aug>
    <au><snm>Ramazanov</snm><fnm>T S</fnm></au>
    <au><snm>Moldabekov</snm><fnm>ZA</fnm></au>
    <au><snm>Gabdullin</snm><fnm>M T</fnm></au>
  </aug>
  <source>Phys Rev E</source>
  <pubdate>2016</pubdate>
  <volume>93</volume>
  <issue>5</issue>
  <fpage>053204</fpage>
</bibl>

<bibl id="B46">
  <title><p>{Ionic transport in high-energy-density matter}</p></title>
  <aug>
    <au><snm>Stanton</snm><fnm>L G</fnm></au>
    <au><snm>Murillo</snm><fnm>M S</fnm></au>
  </aug>
  <source>Physical Review E</source>
  <pubdate>2016</pubdate>
  <volume>93</volume>
  <issue>4</issue>
  <fpage>185</fpage>
</bibl>

<bibl id="B47">
  <title><p>Rydberg Atoms</p></title>
  <aug>
    <au><snm>Gallagher</snm><fnm>T F</fnm></au>
  </aug>
  <publisher>Cambridge University Press</publisher>
  <pubdate>2005</pubdate>
</bibl>

<bibl id="B48">
  <title><p>On the evolution of the phase-space distributions of a
  non-spherical molecular ultracold plasma in supersonic beam</p></title>
  <aug>
    <au><snm>Schulz Weiling</snm><fnm>M</fnm></au>
    <au><snm>Sadeghi</snm><fnm>H</fnm></au>
    <au><snm>Hung</snm><fnm>J</fnm></au>
    <au><snm>Grant</snm><fnm>E R</fnm></au>
  </aug>
  <source>J Phys B</source>
  <pubdate>2016</pubdate>
  <volume>49</volume>
  <fpage>193001</fpage>
</bibl>

<bibl id="B49">
  <title><p>{An experimental and theoretical guide to strongly interacting
  Rydberg gases}</p></title>
  <aug>
    <au><snm>L{\"o}w</snm><fnm>R</fnm></au>
    <au><snm>Weimer</snm><fnm>H</fnm></au>
    <au><snm>Nipper</snm><fnm>J</fnm></au>
    <au><snm>Balewski</snm><fnm>JB</fnm></au>
    <au><snm>Butscher</snm><fnm>B</fnm></au>
    <au><snm>B{\"u}chler</snm><fnm>HP</fnm></au>
    <au><snm>Pfau</snm><fnm>T</fnm></au>
  </aug>
  <source>J Phys B</source>
  <pubdate>2012</pubdate>
  <volume>45</volume>
  <issue>11</issue>
  <fpage>113001</fpage>
</bibl>

<bibl id="B50">
  <title><p>{Long-range interactions between rubidium and potasium Rydberg
  atoms}</p></title>
  <aug>
    <au><snm>Samboy</snm><fnm>N</fnm></au>
  </aug>
  <source>Phys. Rev. A</source>
  <pubdate>2017</pubdate>
  <volume>95</volume>
  <fpage>032702</fpage>
</bibl>

<bibl id="B51">
  <title><p>{Calculation of Rydberg interaction potentials}</p></title>
  <aug>
    <au><cnm>{Sebastian Weber and Christoph Tresp and Henri Menke and Alban
  Urvoy and Ofer Firstenberg and Hans Peter B\"uchler and Sebastian
  Hofferberth}</cnm></au>
  </aug>
  <source>J Phys B</source>
  <pubdate>2017</pubdate>
  <volume>50</volume>
  <issue>13</issue>
  <fpage>133001</fpage>
</bibl>

<bibl id="B52">
  <title><p>Condensed matter theory of dipolar quantum gases</p></title>
  <aug>
    <au><snm>Baranov</snm><fnm>MA</fnm></au>
    <au><snm>Dalmonte</snm><fnm>M</fnm></au>
    <au><snm>Pupillo</snm><fnm>G</fnm></au>
    <au><snm>Zoller</snm><fnm>P</fnm></au>
  </aug>
  <source>Chemical Reviews</source>
  <publisher>ACS Publications</publisher>
  <pubdate>2012</pubdate>
  <volume>112</volume>
  <issue>9</issue>
  <fpage>5012</fpage>
  <lpage>-5061</lpage>
</bibl>

<bibl id="B53">
  <title><p>Cold molecules: theory, experiment, applications</p></title>
  <aug>
    <au><snm>Krems</snm><fnm>R</fnm></au>
    <au><snm>Friedrich</snm><fnm>B</fnm></au>
    <au><snm>Stwalley</snm><fnm>WC</fnm></au>
  </aug>
  <publisher>CRC press</publisher>
  <pubdate>2009</pubdate>
</bibl>

<bibl id="B54">
  <title><p>{Dipole blockade in a cold Rydberg atomic sample
  [Invited]}</p></title>
  <aug>
    <au><snm>Pillet</snm><fnm>P</fnm></au>
    <au><snm>Comparat</snm><fnm>D</fnm></au>
  </aug>
  <source>J Opt Soc Am B</source>
  <pubdate>2010</pubdate>
  <volume>27</volume>
  <issue>6</issue>
  <fpage>A208</fpage>
  <lpage>-A232</lpage>
</bibl>

<bibl id="B55">
  <title><p>{Nonlinear quantum optics mediated by Rydberg
  interactions}</p></title>
  <aug>
    <au><snm>Firstenberg</snm><fnm>O</fnm></au>
    <au><snm>Adams</snm><fnm>C S</fnm></au>
    <au><snm>Hofferberth</snm><fnm>S</fnm></au>
  </aug>
  <source>J Phys B</source>
  <pubdate>2016</pubdate>
  <volume>49</volume>
  <issue>15</issue>
  <fpage>152003</fpage>
</bibl>

<bibl id="B56">
  <title><p>Survival Probabilities in Coherent Exciton Transfer with
  Trapping</p></title>
  <aug>
    <au><snm>M\"ulken</snm><fnm>O</fnm></au>
    <au><snm>Blumen</snm><fnm>A</fnm></au>
    <au><snm>Amthor</snm><fnm>T</fnm></au>
    <au><snm>Giese</snm><fnm>C</fnm></au>
    <au><snm>Reetz Lamour</snm><fnm>M</fnm></au>
    <au><snm>Weidem\"uller</snm><fnm>M</fnm></au>
  </aug>
  <source>Phys. Rev. Lett.</source>
  <publisher>American Physical Society</publisher>
  <pubdate>2007</pubdate>
  <volume>99</volume>
  <fpage>090601</fpage>
</bibl>

<bibl id="B57">
  <title><p>Observing the Dynamics of Dipole-Mediated Energy Transport by
  Interaction-Enhanced Imaging</p></title>
  <aug>
    <au><snm>G{\"u}nter</snm><fnm>G.</fnm></au>
    <au><snm>Schempp</snm><fnm>H.</fnm></au>
    <au><snm>Saint Vincent</snm><fnm>M.</fnm></au>
    <au><snm>Gavryusev</snm><fnm>V.</fnm></au>
    <au><snm>Helmrich</snm><fnm>S.</fnm></au>
    <au><snm>Hofmann</snm><fnm>C. S.</fnm></au>
    <au><snm>Whitlock</snm><fnm>S.</fnm></au>
    <au><snm>Weidem{\"u}ller</snm><fnm>M.</fnm></au>
  </aug>
  <source>Science</source>
  <publisher>American Association for the Advancement of Science</publisher>
  <pubdate>2013</pubdate>
  <volume>342</volume>
  <issue>6161</issue>
  <fpage>954</fpage>
  <lpage>-956</lpage>
</bibl>

<bibl id="B58">
  <title><p>{Coherent Excitation Transfer in a Spin Chain of Three Rydberg
  Atoms}</p></title>
  <aug>
    <au><snm>Barredo</snm><fnm>D</fnm></au>
    <au><snm>Labuhn</snm><fnm>H</fnm></au>
    <au><snm>Ravets</snm><fnm>S</fnm></au>
    <au><snm>Lahaye</snm><fnm>T</fnm></au>
    <au><snm>Browaeys</snm><fnm>A</fnm></au>
    <au><snm>Adams</snm><fnm>CS</fnm></au>
  </aug>
  <source>Phys. Rev. Lett.</source>
  <pubdate>2015</pubdate>
  <volume>114</volume>
  <issue>11</issue>
  <fpage>113002</fpage>
</bibl>

<bibl id="B59">
  <title><p>Coherent many-body spin dynamics in a long-range interacting Ising
  chain</p></title>
  <aug>
    <au><snm>Zeiher</snm><fnm>J</fnm></au>
    <au><snm>Choi</snm><fnm>JY</fnm></au>
    <au><snm>Rubio Abadal</snm><fnm>A</fnm></au>
    <au><snm>Pohl</snm><fnm>T</fnm></au>
    <au><snm>Bijnen</snm><fnm>R</fnm></au>
    <au><snm>Bloch</snm><fnm>I</fnm></au>
    <au><snm>Gross</snm><fnm>C</fnm></au>
  </aug>
  <source>Phys Rev X</source>
  <pubdate>2017</pubdate>
  <volume>7</volume>
  <fpage>041063</fpage>
</bibl>

<bibl id="B60">
  <title><p>Quantum theory of many-particle systems</p></title>
  <aug>
    <au><snm>Fetter</snm><fnm>AL</fnm></au>
    <au><snm>Walecka</snm><fnm>JD</fnm></au>
  </aug>
  <publisher>Courier Corporation</publisher>
  <pubdate>2012</pubdate>
</bibl>

<bibl id="B61">
  <title><p>Quantum magnetism with polar alkali-metal dimers</p></title>
  <aug>
    <au><snm>Gorshkov</snm><fnm>AV</fnm></au>
    <au><snm>Manmana</snm><fnm>SR</fnm></au>
    <au><snm>Chen</snm><fnm>G</fnm></au>
    <au><snm>Demler</snm><fnm>E</fnm></au>
    <au><snm>Lukin</snm><fnm>MD</fnm></au>
    <au><snm>Rey</snm><fnm>AM</fnm></au>
  </aug>
  <source>Phys. Rev. A</source>
  <publisher>American Physical Society</publisher>
  <pubdate>2011</pubdate>
  <volume>84</volume>
  <fpage>033619</fpage>
</bibl>

<bibl id="B62">
  <title><p>Tunable exciton interactions in optical lattices with polar
  molecules</p></title>
  <aug>
    <au><snm>Xiang</snm><fnm>P</fnm></au>
    <au><snm>Litinskaya</snm><fnm>M</fnm></au>
    <au><snm>Krems</snm><fnm>RV</fnm></au>
  </aug>
  <source>Phys. Rev. A</source>
  <publisher>American Physical Society</publisher>
  <pubdate>2012</pubdate>
  <volume>85</volume>
  <fpage>061401</fpage>
</bibl>

<bibl id="B63">
  <title><p>Excitations in organic solids</p></title>
  <aug>
    <au><snm>Agranovich</snm><fnm>VM</fnm></au>
  </aug>
  <publisher>Oxford: Oxford University Press</publisher>
  <pubdate>2009</pubdate>
  <volume>142</volume>
</bibl>

<bibl id="B64">
  <title><p>Rotational spectroscopy of diatomic molecules</p></title>
  <aug>
    <au><snm>Brown</snm><fnm>JM</fnm></au>
    <au><snm>Carrington</snm><fnm>A</fnm></au>
  </aug>
  <publisher>Cambridge University Press</publisher>
  <pubdate>2003</pubdate>
</bibl>

<bibl id="B65">
  <title><p>{Observation of collective excitation of two individual atoms in
  the Rydberg blockade regime}</p></title>
  <aug>
    <au><snm>Ga{\"e}tan</snm><fnm>A</fnm></au>
    <au><snm>Miroshnychenko</snm><fnm>Y</fnm></au>
    <au><snm>Wilk</snm><fnm>T</fnm></au>
    <au><snm>Chotia</snm><fnm>A</fnm></au>
    <au><snm>Viteau</snm><fnm>M</fnm></au>
    <au><snm>Comparat</snm><fnm>D</fnm></au>
    <au><snm>Pillet</snm><fnm>P</fnm></au>
    <au><snm>Browaeys</snm><fnm>A</fnm></au>
    <au><snm>Grangier</snm><fnm>P</fnm></au>
  </aug>
  <source>Nature Physics</source>
  <pubdate>2009</pubdate>
  <volume>5</volume>
  <issue>2</issue>
  <fpage>115</fpage>
  <lpage>-118</lpage>
</bibl>

<bibl id="B66">
  <title><p>Localization in a random XY model with long-range interactions:
  Intermediate case between single-particle and many-body problems</p></title>
  <aug>
    <au><snm>Burin</snm><fnm>AL</fnm></au>
  </aug>
  <source>Phys. Rev. B</source>
  <publisher>American Physical Society</publisher>
  <pubdate>2015</pubdate>
  <volume>92</volume>
  <fpage>104428</fpage>
</bibl>

<bibl id="B67">
  <title><p>Perturbation theory approaches to Anderson and Many-Body
  Localization: some lecture notes</p></title>
  <aug>
    <au><snm>Scardicchio</snm><fnm>A</fnm></au>
    <au><snm>Thiery</snm><fnm>T</fnm></au>
  </aug>
  <source>arXiv preprint arXiv:1710.01234</source>
  <pubdate>2017</pubdate>
</bibl>

<bibl id="B68">
  <title><p>Absence of Diffusion in Certain Random Lattices</p></title>
  <aug>
    <au><snm>Anderson</snm><fnm>P. W.</fnm></au>
  </aug>
  <source>Phys. Rev.</source>
  <publisher>American Physical Society</publisher>
  <pubdate>1958</pubdate>
  <volume>109</volume>
  <fpage>1492</fpage>
  <lpage>-1505</lpage>
</bibl>

<bibl id="B69">
  <title><p>Critical Hamiltonians with long range hopping</p></title>
  <aug>
    <au><snm>Levitov</snm><fnm>L.S.</fnm></au>
  </aug>
  <source>Ann. Phys.</source>
  <publisher>WILEY-VCH Verlag</publisher>
  <pubdate>1999</pubdate>
  <volume>8</volume>
  <issue>7-9</issue>
  <fpage>697</fpage>
  <lpage>-706</lpage>
</bibl>

<bibl id="B70">
  <title><p>Quantum Levy Flights and Multifractality of Dipolar Excitations in
  a Random System</p></title>
  <aug>
    <au><snm>Deng</snm><fnm>X.</fnm></au>
    <au><snm>Altshuler</snm><fnm>B. L.</fnm></au>
    <au><snm>Shlyapnikov</snm><fnm>G. V.</fnm></au>
    <au><snm>Santos</snm><fnm>L.</fnm></au>
  </aug>
  <source>Phys. Rev. Lett.</source>
  <publisher>American Physical Society</publisher>
  <pubdate>2016</pubdate>
  <volume>117</volume>
  <fpage>020401</fpage>
</bibl>

<bibl id="B71">
  <title><p>Duality in Power-Law Localization in Disordered One-Dimensional
  Systems</p></title>
  <aug>
    <au><snm>Deng</snm><fnm>X.</fnm></au>
    <au><snm>Kravtsov</snm><fnm>V. E.</fnm></au>
    <au><snm>Shlyapnikov</snm><fnm>G. V.</fnm></au>
    <au><snm>Santos</snm><fnm>L.</fnm></au>
  </aug>
  <source>Phys. Rev. Lett.</source>
  <publisher>American Physical Society</publisher>
  <pubdate>2018</pubdate>
  <volume>120</volume>
  <fpage>110602</fpage>
</bibl>

<bibl id="B72">
  <title><p>Energy delocalization in strongly disordered systems induced by the
  long-range many-body interaction</p></title>
  <aug>
    <au><snm>Burin</snm><fnm>AL</fnm></au>
  </aug>
  <source>arXiv:cond-mat/0611387</source>
  <pubdate>2006</pubdate>
</bibl>

<bibl id="B73">
  <title><p>Many-Body Localization in Dipolar Systems</p></title>
  <aug>
    <au><snm>Yao</snm><fnm>N. Y.</fnm></au>
    <au><snm>Laumann</snm><fnm>C. R.</fnm></au>
    <au><snm>Gopalakrishnan</snm><fnm>S.</fnm></au>
    <au><snm>Knap</snm><fnm>M.</fnm></au>
    <au><snm>M\"uller</snm><fnm>M.</fnm></au>
    <au><snm>Demler</snm><fnm>E. A.</fnm></au>
    <au><snm>Lukin</snm><fnm>M. D.</fnm></au>
  </aug>
  <source>Phys. Rev. Lett.</source>
  <publisher>American Physical Society</publisher>
  <pubdate>2014</pubdate>
  <volume>113</volume>
  <fpage>243002</fpage>
</bibl>

<bibl id="B74">
  <title><p>{Van der Waals interactions among alkali Rydberg atoms with
  excitonic states}</p></title>
  <aug>
    <au><snm>Zoubi</snm><fnm>H</fnm></au>
  </aug>
  <source>J. Phys. B</source>
  <pubdate>2015</pubdate>
  <volume>48</volume>
  <issue>18</issue>
  <fpage>185002</fpage>
</bibl>

<bibl id="B75">
  <title><p>Many body localization with long range interactions</p></title>
  <aug>
    <au><snm>Nandkishore</snm><fnm>RM</fnm></au>
    <au><snm>Sondhi</snm><fnm>SL</fnm></au>
  </aug>
  <source>Phys Rev X</source>
  <pubdate>2017</pubdate>
  <volume>7</volume>
  <fpage>041021</fpage>
</bibl>

<bibl id="B76">
  <title><p>{Quantum dynamics of disordered spin chains with power-law
  interactions}</p></title>
  <aug>
    <au><snm>Safavi Naini</snm><fnm>A</fnm></au>
    <au><snm>Wall</snm><fnm>M L</fnm></au>
    <au><snm>Acevedo</snm><fnm>O L</fnm></au>
    <au><snm>Rey</snm><fnm>A M</fnm></au>
    <au><snm>Nandkishore</snm><fnm>R M</fnm></au>
  </aug>
  <source>arXiv preprint arXiv:1806.03339</source>
  <pubdate>2018</pubdate>
</bibl>

<bibl id="B77">
  <title><p>Low-frequency conductivity in many-body localized
  systems</p></title>
  <aug>
    <au><snm>Gopalakrishnan</snm><fnm>S</fnm></au>
    <au><snm>M\"uller</snm><fnm>M</fnm></au>
    <au><snm>Khemani</snm><fnm>V</fnm></au>
    <au><snm>Knap</snm><fnm>M</fnm></au>
    <au><snm>Demler</snm><fnm>E</fnm></au>
    <au><snm>Huse</snm><fnm>DA</fnm></au>
  </aug>
  <source>Phys. Rev. B</source>
  <publisher>American Physical Society</publisher>
  <pubdate>2015</pubdate>
  <volume>92</volume>
  <fpage>104202</fpage>
</bibl>

<bibl id="B78">
  <title><p>Absence of many-body mobility edges</p></title>
  <aug>
    <au><snm>De Roeck</snm><fnm>W</fnm></au>
    <au><snm>Huveneers</snm><fnm>F</fnm></au>
    <au><snm>M\"uller</snm><fnm>M</fnm></au>
    <au><snm>Schiulaz</snm><fnm>M</fnm></au>
  </aug>
  <source>Phys. Rev. B</source>
  <publisher>American Physical Society</publisher>
  <pubdate>2016</pubdate>
  <volume>93</volume>
  <fpage>014203</fpage>
</bibl>

<bibl id="B79">
  <title><p>Stability and instability towards delocalization in many-body
  localization systems</p></title>
  <aug>
    <au><snm>De Roeck</snm><fnm>W</fnm></au>
    <au><snm>Huveneers</snm><fnm>{Fran\ifmmode \mbox{\c{c}}\else
  \c{c}\fi{}ois}</fnm></au>
  </aug>
  <source>Phys. Rev. B</source>
  <publisher>American Physical Society</publisher>
  <pubdate>2017</pubdate>
  <volume>95</volume>
  <fpage>155129</fpage>
</bibl>

<bibl id="B80">
  <title><p>Rare-region effects and dynamics near the many-body localization
  transition</p></title>
  <aug>
    <au><snm>Agarwal</snm><fnm>K</fnm></au>
    <au><snm>Altman</snm><fnm>E</fnm></au>
    <au><snm>Demler</snm><fnm>E</fnm></au>
    <au><snm>Gopalakrishnan</snm><fnm>S</fnm></au>
    <au><snm>Huse</snm><fnm>DA</fnm></au>
    <au><snm>Knap</snm><fnm>M</fnm></au>
  </aug>
  <source>Ann. Phys.</source>
  <pubdate>2017</pubdate>
  <volume>529</volume>
  <issue>7</issue>
  <fpage>1600326</fpage>
  <lpage>-n/a</lpage>
  <note>1600326</note>
</bibl>

<bibl id="B81">
  <title><p>{Thermal inclusions: how one spin can destroy a many-body localized
  phase.}</p></title>
  <aug>
    <au><snm>Ponte</snm><fnm>P</fnm></au>
    <au><snm>Laumann</snm><fnm>C R</fnm></au>
    <au><snm>Huse</snm><fnm>DA</fnm></au>
    <au><snm>Chandran</snm><fnm>A</fnm></au>
  </aug>
  <source>Phil. Trans. R. Soc. A</source>
  <pubdate>2017</pubdate>
  <volume>375</volume>
  <issue>2108</issue>
  <fpage>20160428</fpage>
</bibl>

<bibl id="B82">
  <title><p>On the stability of many-body localization in $ d> 1$</p></title>
  <aug>
    <au><snm>Potirniche</snm><fnm>ID</fnm></au>
    <au><snm>Banerjee</snm><fnm>S</fnm></au>
    <au><snm>Altman</snm><fnm>E</fnm></au>
  </aug>
  <source>arXiv preprint arXiv:1805.01475</source>
  <pubdate>2018</pubdate>
</bibl>

<bibl id="B83">
  <title><p>{Decay Dynamics of the Predissociating High Rydberg States of
  NO}</p></title>
  <aug>
    <au><snm>Remacle</snm><fnm>F</fnm></au>
    <au><snm>Vrakking</snm><fnm>M</fnm></au>
  </aug>
  <source>J Phys Chem A</source>
  <pubdate>1998</pubdate>
  <volume>102</volume>
  <fpage>9507</fpage>
  <lpage>9517</lpage>
</bibl>

</refgrp>
} 

\enddocument